\documentclass[11pt]{article}

\usepackage[cmtip,arrow]{xy}
\usepackage{pb-diagram,pb-xy}

\newlength{\vshift}
\newlength{\hshift}
\usepackage{amssymb,amsopn}

\def\la{\lambda}

\def\ka{\kappa}

\def\si{\sigma}

\def\ds{\stackrel{\star}{,}}

\def\p{\partial}

\def\h{\hat}
\def\lb{\lbrack}
\def\rb{\rbrack}
\def\p{\partial}

\begin{document}

\begin{titlepage}
\rightline{LMU-ASC 32/05}
\rightline{MPP-2005-27}
\rightline{IUB-TH-056}
\rightline{hep-th/0504183}

%\vspace{1.0em}
\begin{center}

{\LARGE{\bf A Gravity Theory on Noncommutative Spaces}}

\vskip 3em

{{\bf Paolo Aschieri${}^{5}$, Christian Blohmann${}^{3,6}$, Marija Dimitrijevi\' c${}^{1,2,4}$, \\
Frank Meyer${}^{1,2}$, Peter Schupp${}^{3}$ and Julius Wess${}^{1,2}$ }}

\vskip 1em

${}^{1}$Arnold Sommerfeld Center for Theoretical Physics\\
Universit\"at M\"unchen, Fakult\"at f\"ur Physik\\
Theresienstr.\ 37, 80333 M\"unchen, Germany\\[1em]

${}^{2}$Max-Planck-Institut f\"ur Physik\\
F\"ohringer Ring 6, 80805 M\"unchen, Germany\\[1em]

${}^{3}$International University Bremen,\\ 
Campus Ring 8, 28759 Bremen, Germany\\[1em]

${}^{4}$University of Belgrade, Faculty of Physics\\
Studentski trg 12, 11000 Beograd, Serbia and Montenegro\\[1em]

${}^{5}$Dipartimento di Scienze e Tecnologie Avanzate\\
Universit\' a del Piemonte Orientale, and INFN\\
Via Bellini 25/G 15100 Alessandria, Italy\\[1em]

${}^{6}$University of California, Department of Mathematics\\
Berkeley, California 94720-3840\\[1em]

\end{center}

%\vspace*{0.5cm}

\begin{abstract}
A deformation of the algebra of diffeomorphisms is constructed for canonically deformed spaces with
constant deformation parameter $\theta$. The algebraic relations remain the same, whereas the comultiplication rule
(Leibniz rule) is different from the undeformed one. Based on this deformed algebra a covariant
tensor calculus is constructed and all the concepts like metric, covariant derivatives, curvature
and torsion can be defined on the deformed space as well. The construction of these geometric
quantities is presented in detail. This leads to an action invariant under the deformed
diffeomorphism algebra  and can be interpreted as a $\theta$-deformed Einstein-Hilbert action. The metric or the vierbein field will be the
dynamical variable as they are in the undeformed theory. The action and all relevant quantities are expanded up to second order in $\theta$.

\end{abstract}
%\vspace*{0.2cm}

\scriptsize{PACS numbers: 02.40.Gh, 02.20.Uw, 04.20.-q, 04.60.-m, 11.10.Nx \\ \\
E-mail: aschieri,dmarija,meyerf,wess@theorie.physik.uni-muenchen.de;
blohmann@math.berkeley.edu; \\ \hspace{0.5cm} p.schupp@iu-bremen.de}
\vfill

\end{titlepage}\vskip.2cm

\newpage
\setcounter{page}{1}
\newcommand{\Section}[1]{\setcounter{equation}{0}\section{#1}}
\renewcommand{\theequation}{\arabic{section}.\arabic{equation}}

\Section{Introduction}

Several arguments are presently used to motivate a deviation from the flat-space concept at very short
distances \cite{Wheeler:1962,Witten:1962}. Among the new concepts are quantum spaces
\cite{Pauli:1985,Snyder:1947,Manin:1989sz,wess:1991}. They have the advantage that their mathematical structure is well defined and that, based on this structure, questions on the physical behaviour of these systems can be asked. One of the questions is if physics on quantum spaces can be formulated by field equations, how they deviate from the usual field equations, and to what changes they lead in their physical interpretation.

Quantum spaces depend on parameters such that for a particular value of these parameters they become the usual flat
space. Thus, we call them deformed spaces. In the same sense we expect a deformation of the field equations and
finally a deformation of their physical predictions \cite{Behr:2002wx,Ohl:2004tn,Melic:2005am,Martin:2005gt,Grosse:2001xz}.

Several of these deformations have been studied \cite{Madore:2000en,wess:2005}. They are all based on nontrivial commutation
relations of the coordinates. This algebraic deformation leads to a star product formulation as it is used for
deformation quantisation \cite{Bayen:1978ha,Gayral:2005ih,Jambor:2004kc}. In this paper we start from the star
product deformation and consider the algebraic relations as consequences. We might not have started from the most
general realization of the deformed algebra, but certainly from one that is very useful for physical interpretation.
This way deformed gauge theories have been constructed by the use of the Seiberg-Witten map
\cite{Seiberg:1999vs,Jurco:2000ja,Jurco:2001rq,Jurco:2001my,Meyer:2003wj,Dimitrijevic:2003pn,Behr:2003qc,Schupp:2001we}. Their field content is the same as in the undeformed theory, the deformation parameters enter in the deformed field equations as coupling constants.

The question was still open if the gravity theories can be treated the same way and has been investigated by several
authors
\cite{Chamseddine:1992yx,Madore:1992,Castellani:1993ud,Castellani:1994yh,Madore:1996bb,Connes:1996gi,Chamseddine:2000si,Moffat:2000gr,Vacaru:2000yk,Cardella:2002pb,Cacciatori:2002gq,Cacciatori:2002ib,Ardalan:2002qt,Garcia-Compean:2002ss,Garcia-Compean:2003pm}.
We present here a positive answer to this question based on a deformed algebra of diffeomorphisms and this way avoiding the concept of
general coordinate transformations. In this presentation we restrict ourselves to the discussion of the canonical quantum space with
$\theta^{\mu\nu}$ constant. The construction is now not based on Seiberg-Witten
maps. In a forthcoming paper we shall show how this can be generalized to $x$-dependent $\theta^{\mu\nu}$.

By outlining the content of the individual sections we will show the strategy by which a deformed gravity theory can be constructed.

In section 2 we give a short introduction to the $\theta$-deformed quantum algebra defined by the Moyal-Weyl product. Emphasis is on those concepts that shall be used in the rest of the paper. More detailed features of this algebra can be found in the literature and we give some relevant references.

In section 3 the concept of derivatives is introduced. It turns out that there is a natural way to define a derivative on the quantum algebra. We investigate these derivatives as elements of a Hopf algebra and find that the usual derivatives and the derivatives on the quantum space represent the same Hopf algebra.

We also generalize the derivatives to higher order differential operators and define algebras of higher order differential operators both acting on differential manifolds and acting on the deformed space.
A map from the algebra of functions on the differential manifold to the algebra of functions on the deformed space is constructed. This map will be the basis for the representation of the diffeomorphism algebra by an algebra of higher order differential operators acting on the deformed space.

In section 4 we study the algebra generated by vector fields and exhibit its Hopf algebra structure. It is the algebra of diffeomorphisms derived from general coordinate transformations. Scalar, vector, and  tensor fields are representations of this algebra.

In section 5 we construct a morphism between the classical algebra of diffeomorphisms and an algebra acting on the deformed space. At first, this is an algebra morphism but not a Hopf algebra morphism. To find a comultiplication rule we derive the Leibniz rule for the deformed algebra and show that it can be obtained from an abstract comultiplication which we construct explicitly to all orders in $\theta$. Thus, we have constructed a new Hopf algebra of diffeomorphisms as a deformation of the classical one. A deformed gravity theory will now be investigated as a theory covariant under this deformed Hopf algebra.

In section 6 we restrict the formalism developed so far to vector fields linear in the coordinates. They form a subalgebra. The Lorentz algebra can be obtained in that way and we find a representation of the Lorentz algebra
by differential operators that act on the deformed space. The comultiplication rule follows from the general formalism and shows that the derivatives have to be part of the algebra. This way we have found a representation of the Poincar\' e algebra with nontrivial comultiplication rule. A tensor calculus of fields is developed for this algebra and invariant actions are constructed. All the operations in the definition of the Lagrangian --- derivatives and multiplication --- are in the deformed algebra. Field equations can be obtained that are Lorentz covariant. This by itself is an interesting result but it also serves as a guideline for the construction of a general theory on the deformed space.

In section 7 we show that all the concepts of differential geometry like tensor fields, covariant derivatives, connection and curvature can be obtained by a map from the usual commutative space to the deformed space. The relevant formulas are calculated explicitly.

In section 8 and 9 we turn to a metric space. We define the metric as a symmetric and real tensor that coincides with $g_{\mu\nu}$ in the limit $\theta\to 0$. All other geometrical quantities are constructed in terms of this metric. Finally, we use the curvature scalar expressed in terms of $g_{\mu\nu}$ to construct a Lagrangian for a deformed gravity theory.

In section 10 we expand all these quantities up to second order in $\theta$. The action obtained this way can be used to calculate some effects of the deformation. The deformation parameter $\theta$ enters as a coupling constant as it is familiar from gauge theory.

This way it is possible to study deviations from the undeformed classical gravity due to space-time noncommutativity. The strategy developed
here can  be generalized to other $\star$-products which then lead to other algebraic structures of space-time.

\Section{$\theta$-deformed coordinate algebra}

A simple example of a noncommutative coordinate algebra is the $\theta$-deformed or canonical quantum algebra
${\cal{A}}_\theta$. It is based on the relations \cite{Chu:1998qz,Schomerus:1999ug}
\begin{equation}
\lb \h{x}^\mu , \h{x}^\nu\rb = i \theta^{\mu\nu}, \label{2.1}
\end{equation}
with $\theta^{\mu\nu}$ constant and real.

This algebra can be realized on the linear space ${\cal{F}}$ of complex functions $f(x)$ of commuting variables: The elements of the algebra ${\cal{A}}_\theta$ are represented by functions of the commuting variables
$f(x)$, their product by the Moyal-Weyl star-product ($\star$-product) \cite{Moyal:1949sk,Bayen:1978ha}
\begin{equation}
f\star g\,(x) = e^{\frac{i}{2}\frac{\p }{\p x^\rho}
\theta^{\rho\sigma}\frac{\p }{\p y^\sigma}}f(x)g(y)
\Big|_{y\rightarrow x} .\label{2.2}
\end{equation}
This $\star$-product of two functions is a function again. The $\star$-product defines the associative but noncommutative algebra ${\cal{A}}_\theta$.  By taking the usual pointwise product of two functions we obtain the usual algebra of functions. This algebra is associative and commutative. We shall call it ${\cal{A}}_f$. Note that we write $f(x)$ for elements of ${\cal{A}}_f$ as well as for elements of ${\cal{A}}_\theta$.

As far as complex conjugation is concerned we observe that the $\star$-product of two real functions is not real again. Denoting the complex conjugate of $f$ by $\bar{f}$ we find from definition (\ref{2.2})
\begin{equation}
\overline{f\star g} = \bar{g}\star\bar{f} .\label{2.3}
\end{equation}
%Defining the transposed of the $\star$-product by
%\begin{equation}
%(f\star g)^{{\mbox{\small T}}} =g\star f \label{2.4}
%\end{equation}
%we obtain after an obvious definition of hermitian conjugate ($+$)
%\begin{equation}
%{\overline{f\star g}}^{{\mbox{\small T}}}  = (f\star g)^+ = \bar{f}\star\bar{g} .\label{2.5}
%\end{equation}
%For real functions $f$ and $g$ $f\star g$ is hermitian.
{}From definition (\ref{2.2}) it also follows that
\begin{equation}
x^\mu\star x^\nu -x^\nu\star x^\mu =i\theta^{\mu\nu} .\label{2.4}
\end{equation}
These are the defining relations for the generators of the algebra ${\cal{A}}_\theta$. Any element of the space of ordinary functions represents an element of ${\cal{A}}_\theta$; there is an invertible
map $\phi$ \cite{Waldmann:thesis}
\begin{equation}
\phi:\> {\cal F} \to {\cal{A}}_\theta .\label{2.5}
\end{equation}

If we know the elements that are represented by the functions $f$ and $g$ we can ask how the pointwise product of two functions  $f\cdot g$ is represented in ${\cal{A}}_\theta$ by  $\star$-products of $f$ and $g$ and their derivatives. First an example
\begin{equation}
x^\mu\cdot x^\nu= x^\mu\star x^\nu -\frac{i}{2}\hspace*{0.9mm} \theta^{\mu\nu}.\label{2.6}
\end{equation}
This follows from (\ref{2.2}). The pointwise product $x^\mu\cdot x^\nu$ as an element of ${\cal{A}}_\theta$ represents the sum of two elements of ${\cal{A}}_\theta$ modulo the relations (\ref{2.4}). In general $f\cdot g$ will represent a sum of  $\star$-products of $f$, $g$ and their derivatives
\begin{equation}
f\cdot g = \sum _{n=0}^\infty \Big(-\frac{i}{2}\Big)^n
\frac{1}{n!}\theta^{\rho_1\sigma_1}\dots
\theta^{\rho_n\sigma_n}\Big(\partial_{\rho_1}\dots\partial_{\rho_n}f\Big) \star
\Big(\partial_{\sigma_1}\dots\partial_{\sigma_n}g\Big) .\label{2.7}
\end{equation}
This is a well defined formula because the derivatives of functions are functions again and we know how to $\star$-multiply them. Applied to $x^\mu x^\nu$ the equation (\ref{2.7}) reproduces (\ref{2.6}). The operations on the right hand side of (\ref{2.7}) are all in ${\cal{A}}_\theta$. To prove (\ref{2.7}) we use the $\star$-product in the form that makes use of the tensor product of the vector spaces ${\cal F}$
\begin{equation}
f\star g = \mu \{ e^{\frac{i}{2}\theta^{\rho\sigma}\p_\rho \otimes\p_\sigma}f\otimes g \}.
\label{2.8} 
\end{equation}
The bilinear map $\mu$ maps the tensor product to the space of functions
\begin{eqnarray}
\mu : {\cal F}\otimes {\cal F} \hspace*{-2mm}&\to&\hspace*{-2mm} {\cal F} \nonumber\\
\mu\{ f\otimes g\} \hspace*{-2mm}&\mapsto&\hspace*{-2mm} f\cdot g .\label{2.9}
\end{eqnarray}
We now use the obvious equation
\begin{equation}
f\cdot g= \mu \{ e^{\frac{i}{2}\theta^{\rho\sigma}\p_\rho \otimes\p_\sigma}
e^{-\frac{i}{2}\theta^{\rho\sigma}\p_\rho \otimes\p_\sigma} f\otimes g \}.
\label{2.10} 
\end{equation}
The first exponent will produce $\star$-products, the second one the sum of terms in (\ref{2.7}). 
On the other hand, equation (\ref{2.2}) expresses the $\star$-product $f\star g$ in terms of pointwise products of $f$ and $g$ and their derivatives. All these operations are in ${\cal{A}}_f$.  

\Section{Derivatives on ${\cal{A}}_\theta$}

Derivatives on quantum spaces were constructed in \cite{Woronowicz:1989rt,Wess:1991vh}.
There is, however, a natural way to introduce derivatives on ${\cal{A}}_\theta$ based on the $\star$-product formulation. We know that the derivative of a function $f\in {\cal F}$ is a function again. This function can be mapped to ${\cal{A}}_\theta$, the image we call the $\star$-derivative of $f\in {\cal{A}}_\theta$
\begin{equation}
%\label{eq:Diagram1}
\begin{diagram}
  \node{f\in {\cal F}} 
  \arrow{e,t,T}{\phi} \arrow{s,l,T}{\partial_\mu}
  \node{f \in {\cal{A}}_\theta} 
  \arrow{s,r,T}{\partial^\star_\mu}
  \\      
  \node{(\partial _\mu f) \in {\cal F}} 
  \arrow{e,t,T}{\phi}
  \node{(\partial^\star _\mu\triangleright f) \in {\cal{A}}_\theta}
\end{diagram}
\end{equation}
% \begin{eqnarray}
% {f\in {\cal F}}&\stackrel{\phi}{\longmapsto}& {f \in {\cal{A}}_\theta} \nonumber\\
% \partial_\mu\downarrow &&\downarrow\partial^\star_\mu \nonumber\\
% {(\partial _\mu f) \in {\cal F}}&\stackrel{\phi}{\longmapsto}& {(\partial^\star _\mu\triangleright f) \in {\cal{A}}_\theta}.\nonumber
% \end{eqnarray}
This defines $\partial_\mu^\star$ acting on $f \in {\cal{A}}_\theta$
\begin{equation}
\p^\star_\mu\triangleright f := (\p_\mu f) .\label{3.1}
\end{equation}

Now we discuss a few properties of the $\star$-derivatives. From the definition (\ref{3.1}) follows 
\begin{equation}
\p^\star _\mu\triangleright x^\rho =\delta _\mu^\rho ,\label{3.2}
\end{equation}
a property that we demand for a reasonable definition of a derivative. 
As the $\star$-product of two functions is a function again we can use the definition (\ref{3.1}) to differentiate
$f\star g$
\begin{equation}
\p^\star_\mu\triangleright (f\star g) = \left( \p_\mu(f\star g) \right) .\label{3.3}
\end{equation}
For the $\star$-product with $x$-independent $\theta$ it follows from (\ref{2.2}) that
\begin{equation}
(\p_\mu (f\star g)) = (\p_\mu f)\star g + f\star (\p_\mu g) .\label{3.4}
\end{equation}
Using (\ref{3.1}) we obtain
\begin{equation}
\p^\star_\mu\triangleright (f\star g) = (\p^\star_\mu \triangleright f)\star g +f\star (\p^\star_\mu\triangleright g) .\label{3.5}
\end{equation}
In this equation all operations, derivative and product, are within ${\cal{A}}_\theta$. We have expressed the $\star$-derivative acting on a $\star$-product by the $\star$-product of $\star$-derivatives.

Applying this rule to (\ref{2.4}) we find
\begin{equation}
\p_\mu^\star\triangleright \Big( \lb x^\rho \ds x^\sigma\rb - i \theta^{\rho\sigma} \Big) =  0 .\label{3.6}
\end{equation}
This confirms that the derivative (\ref{3.1}) is a well defined map on ${\cal{A}}_\theta$. Moreover, from (\ref{3.1}) follows
\begin{equation}
\p_\mu^\star \triangleright(\p_\nu^\star \triangleright f) = (\p_\mu\p_\nu f) \label{3.7}
\end{equation}
and therefore
\begin{equation}
\p_\mu^\star \triangleright(\p_\nu^\star \triangleright f) 
=\p_\nu^\star \triangleright(\p_\mu^\star \triangleright f) .\label{3.8}
\end{equation}
The action of $\star$-derivatives on a function is commutative.

Derivatives were defined by their action on functions but they can be seen as differential operators as well because equation (\ref{3.5}) holds for any function $g$. Thus, it gives rise to the operator equation
\begin{equation}
\p _\mu^\star \star f = (\p ^\star_\mu \triangleright f)  +f\star \p _\mu^\star .\label{3.9}
\end{equation}
We use the $\star$ when the derivative is meant to be an operator. As for ordinary derivatives, we
can also use the bracket notation if the derivatives act on a function. To emphasize that the
action is meant we also use the triangle notation
\begin{equation}
(\p _\mu^\star \star f) \equiv \p ^\star_\mu \triangleright f .\label{3.10}
\end{equation}

Taking for $f$ the coordinate $x^\rho$ we obtain from (\ref{3.9})
\begin{equation}
\lb \p_\mu^\star \ds x^\rho \rb  = \delta^\rho_\mu  .\label{3.11}
\end{equation}
Analogously to equation (\ref{3.6}) we get
\begin{equation}
\p_\mu^\star \star \Big( \lb x^\rho \ds x^\sigma\rb - i \theta^{\rho\sigma} \Big) 
= \Big( \lb x^\rho \ds x^\sigma\rb - i \theta^{\rho\sigma} \Big) \star \p_\mu^\star   .\label{3.12}
\end{equation}
 Equation (\ref{3.8}), valid for any function $g$, leads to the commutativity of $\star$-derivative operators
 \begin{equation}
 \lb \p^\star _\mu \ds \p^\star _\nu\rb = 0 .\label{3.13}
 \end{equation}

The derivatives, as maps on the algebra ${\cal A}_\theta$ have a Hopf algebra structure
\cite{Abe:1980,Chari-Presley,Klimyk:1997eb}. This implies the following properties: The derivatives generate an algebra with defining relation (\ref{3.13}). The coproduct is defined as follows
\begin{equation}
\Delta(\p_\mu^\star) = \p_\mu^\star \otimes 1  + 1\otimes\p_\mu^\star . \label{3.14}
\end{equation}
It is compatible with the algebra
\begin{equation}
\lb \Delta(\p^\star_\mu) \ds \Delta(\p^\star_\nu)\rb = 0 . \label{3.14'}
\end{equation}
The coassociativity 
\begin{equation}
(\Delta\otimes {\mbox{id}})\circ\Delta = ({\mbox{id}}\otimes \Delta)\circ\Delta \label{3.15}
\end{equation}
can be verified explicitly. When we apply (\ref{3.15}) to $\p_\mu^\star$ we obtain
\begin{eqnarray}
\Big( (\Delta\otimes {\mbox{id}})\circ\Delta \Big) (\p_\mu^\star) \hspace*{-2mm}&=&\hspace*{-2mm}
(\Delta\otimes {\mbox{id}}) (\p_\mu^\star\otimes 1  + 1\otimes\p_\mu^\star) \nonumber\\
&=&\hspace*{-2mm} (\p_\mu^\star\otimes 1  + 1\otimes\p_\mu^\star)\otimes {\mbox{id}} + 
{\mbox{id}}\otimes {\mbox{id}}\otimes \p_\mu^\star . \label{3.16}
\end{eqnarray}
That $({\mbox{id}}\otimes \Delta)\circ\Delta$ gives the same result can be easily seen. To define a Hopf algebra we still need a counit and an antipode. They are given by
\begin{equation}
\varepsilon(\p_\mu^\star) =0, \quad\quad S(\p^\star_\mu)=-\p^\star_\mu  .\label{3.14''}
\end{equation}
The Leibniz rule (\ref{3.5}) can be obtained by applying the bilinear map $\mu_\star \{f\otimes g\} = f\star g$ to the coproduct
\begin{eqnarray}
\mu_\star \{\Delta(\p_\mu^\star) \triangleright f\otimes g\} \hspace*{-2mm}&=&\hspace*{-2mm}
\mu_\star \{(\p_\mu^\star \star f)\otimes g + f\otimes (\p_\mu^\star \star g)\} \nonumber\\
&=&\hspace*{-2mm} (\p_\mu^\star \triangleright f)\star g + f\star (\p_\mu^\star \triangleright g) \label{3.17}
\end{eqnarray}
The usual derivatives $\p_\mu$ and $\star$-derivatives $\p^\star_\mu$ are representations
of the same Hopf algebra.

We are going to discuss the algebra of higher order differential operators. Acting on ${\cal A}_f$, elements of this algebra are
\begin{equation}
D = \sum_r d^{\mu_1\dots\mu_r}(x)\p_{\mu_1}\dots\p_{\mu_r} .\label{3.18}
\end{equation}
Acting on ${\cal A}_\theta$, the elements are
\begin{equation}\label{star-diff-op}
D^\star = \sum_r d^{\mu_1\dots\mu_r}(x)\p^\star_{\mu_1}\dots\p^\star_{\mu_r} \label{3.18'}
\end{equation}
where the coefficient function $d^{\mu_1\dots\mu_r}(x)$ has to be considered as an element of ${\cal A}_\theta$. The multiplication of the operators $D$ is standard. The multiplication of $\star$-operators can be defined if we consider the algebra ${\cal{A}}_\theta$ extended by the derivatives. It is always possible to write such a product in the form as in  (\ref{star-diff-op}) with all derivatives on the right by using the operator equation following from the Leibniz rule. This multiplication can essentially be obtained by replacing the product of the coefficient functions by the $\star$-product.

The operator $D$ can be mapped to operators acting on $\mathcal{A}_\theta$. To construct such a map we reexamine the pointwise product of two functions (\ref{2.7}) in the light of higher order differential operators
\begin{equation}
f\cdot g = X_f^\star \triangleright g = (X^\star _f \star g), \label{3.19}
\end{equation}
where
\begin{equation}
X^\star _f = \sum_{n=0}^\infty \frac{1}{n!}\Big(-\frac{i}{2}\Big)^n
\theta^{\rho_1\sigma_1}\dots\theta^{\rho_n\sigma_n}(\partial_{\rho_1}\dots\partial_{\rho_n}f)
\star \partial^\star _{\sigma_1}\dots\partial^\star _{\sigma_n} .\label{3.20}
\end{equation}
This can easily be generalized to the action of differential operators on $g$
\begin{equation}
(Dg) = X_D^\star \triangleright g ,\label{3.22}
\end{equation}
where
\begin{equation}
X^\star _D = \sum_{n=0}^\infty \frac{1}{n!}\Big(-\frac{i}{2}\Big)^n
\theta^{\rho_1\sigma_1}\dots\theta^{\rho_n\sigma_n}(\partial_{\rho_1}\dots
\partial_{\rho_n}d^{\mu_1\dots\mu_r}(x))
\star \partial^\star _{\sigma_1}\dots\partial^\star _{\sigma_n}\p^\star_{\mu_1}\dots\p^\star_{\mu_r} .\label{3.23}
\end{equation}

Now we define a map
\begin{eqnarray}
f &\mapsto& X^\star_f \nonumber\\
D &\mapsto& X^\star_D .\label{3.23'} 
\end{eqnarray}
This map is actually an algebra map if the multiplications of differential operators are defined as
above after (\ref{star-diff-op}). From (\ref{3.22}) follows
\begin{equation}
(D\cdot D' g) = (X^\star _D \star X^\star _{D'} )\triangleright g .\label{3.24}
\end{equation}
This is true for any function $g$ and thus the map (\ref{3.23'}) can be interpreted as a morphism of algebras.

\Section{Diffeomorphisms}

We will develop a formalism by which the algebra of diffeomorphisms acting on ${\cal A}_f$ can be mapped to an algebra of $\star$-differential operators acting on ${\cal A}_\theta$. 

Let us first recall the concept of diffeomorphisms as a Hopf algebra. They are generated by vector fields 
acting on a differential manifold. The vector fields are defined as follows
\begin{equation}
\xi=\xi^\mu(x)\frac{\p}{\p x^\mu}. \label{4.1}
\end{equation}
The commutator of two vector fields $\xi,\eta$ is again a vector field:
\begin{equation}
\lb \xi, \eta \rb = \xi\times \eta , \label{4.2}
\end{equation}
where the vector field $\xi\times \eta$ is given by
\begin{equation}
\xi\times \eta =\Big( \eta^\mu(\p_\mu \xi^\rho) -\xi^\mu (\p_\mu \eta^\rho) \Big) \frac{\p}{\p x^\rho} .\label{4.3}
\end{equation}
{}From the Leibniz rule for derivatives follows the Leibniz rule for vector fields
\begin{equation}
(\xi (f\cdot g)) = (\xi f)\cdot g +f\cdot (\xi g) . \label{4.4}
\end{equation}
This Leibniz rule follows from an abstract comultiplication rule that defines the action of a vector field on a tensor product
\begin{equation}
\Delta(\xi) = \xi\otimes 1  + 1\otimes\xi .\label{4.5}
\end{equation}
It can be verified with (\ref{4.2}) that the comultiplication (\ref{4.5}) and the algebraic relation are compatible without making use of $\xi$ represented as a differential operator
\begin{equation}
\lb \Delta (\xi), \Delta (\eta) \rb = \Delta (\xi\times \eta) . \label{4.6}
\end{equation}
This defines a bialgebra. With the counit and the antipode
\begin{equation}
\varepsilon (\xi) =0,\quad\quad S(\xi)=-\xi  \label{4.7}
\end{equation}
it becomes a Hopf algebra. Here $\xi$ and $\eta$ need to be treated as abstract objects. Their product $\xi\eta$ is to be viewed as an abstract product modulo the relation $\xi \eta - \eta \xi = \xi \times \eta$\footnote{In other words, we are considering the universal enveloping algebra freely generated by elements $\xi,\eta$ modulo the relation $\xi \eta - \eta \xi = \xi \times \eta$.}. 

Diffeomorphisms are intimately connected with general coordinate transformations defined as follows
\begin{equation}
x^\mu\rightarrow x'^\mu=x^\mu + \xi^\mu(x) \label{4.8}
\end{equation}
with infinitesimal $\xi^\mu(x)$. 

A scalar field is defined to transform under general coordinate transformations as follows
\begin{eqnarray}
\phi'(x') = \phi(x) .\nonumber
\end{eqnarray}
For infinitesimal transformations (\ref{4.8}) this becomes
\begin{equation}
\delta_\xi\phi(x) = \phi'(x)-\phi(x) = -\xi^\mu(\partial_\mu\phi(x)) 
=-(\xi\phi (x)) .\label{4.9}
\end{equation}
Similarly we define covariant vector fields
\begin{equation}
\delta_\xi V_\mu = -\xi^\rho(\partial_\rho V_\mu) -(\partial _\mu \xi^\rho)V_\rho \label{4.10}
\end{equation}
and contravariant vector fields
\begin{equation}
\delta_\xi V^\mu = -\xi^\rho(\partial_\rho V^\mu) +(\partial _\rho \xi^\mu)V^\rho .\label{4.11}
\end{equation}
This can be easily generalized to tensor fields with an arbitrary number of covariant and contravariant indices.

These transformations represent the algebra of diffeomorphisms (\ref{4.2})
\begin{equation}
\lb \delta_\xi,\delta_\eta \rb = \delta_{\xi\times\eta } ,\label{4.12}
\end{equation}
with the coproduct
\begin{equation}
\Delta \delta_\xi = \delta_\xi\otimes 1 + 1\otimes \delta_\xi  .\label{4.13}
\end{equation}

As a consequence of (\ref{4.13}) the product of two vector fields transforms like a tensor field of second rank
\begin{eqnarray}
\delta_\xi (V_\mu V_\nu) = \mu \{\Delta(\delta_\xi) V_\mu \otimes V_\nu \}
& = & \mu \{(\delta_\xi V_\mu) \otimes V_\nu + V_\mu \otimes (\delta_\xi V_\nu)  \}  \label{4.14} \\
& =& -\xi^\rho \partial_\rho (V_\mu V_\nu) -   (\partial_\mu \xi^\rho) (V_\rho V_\nu) -  (\partial_\nu \xi^\rho)(V_\mu V_\rho) . \nonumber
\end{eqnarray}
This can easily be extended to the product of arbitrary tensor fields.

We summarize the Hopf algebra structure of general coordinate transformations
\begin{eqnarray}
\lb \delta_\xi,\delta_\eta \rb \hspace*{-2mm}&=&\hspace*{-2mm} \delta_{\xi\times\eta },
\quad\quad \varepsilon (\delta_\xi) =0, \quad\quad S(\delta_\xi) = -\delta_\xi, \nonumber\\
\Delta \delta_\xi  \hspace*{-2mm}&=&\hspace*{-2mm} \delta_\xi\otimes 1 + 1\otimes \delta_\xi  ,\quad\quad
\lb \Delta(\delta_\xi),\Delta(\delta_\eta) \rb = \Delta(\delta_{\xi\times\eta }) . \label{4.15}
\end{eqnarray}
This is true for any realization of $\delta_\xi$ on arbitrary tensor fields. It is a property of the abstract Hopf algebra and not of a particular representation as differential operator.

\Section{Diffeomorphism algebra on ${\cal A}_\theta$}

We know how to map the algebra of higher order classical differential operators acting on ${\cal A}_f$ into the corresponding algebra acting on ${\cal A}_\theta$. The relevant formulas are (\ref{3.20}) and (\ref{3.23}). 

In the same way, the action of a vector field
\begin{equation}
\xi=\xi^\mu\p_\mu \label{5.1}
\end{equation}
can be mapped to a higher order differential operator acting on ${\cal A}_\theta$
\begin{equation}
(\xi\cdot f) = X^\star_\xi \triangleright f .\label{5.2}
\end{equation}
{}From (\ref{3.24}) then follows
\begin{equation}
\lb X_\xi^\star \ds X^\star_\eta \rb = X^\star _{\xi\times \eta} .\label{5.3}
\end{equation}
The operators $X^\star_\xi$ represent the algebra of vector fields. To obtain a Leibniz rule on ${\cal A}_\theta$ we apply the operator $X^\star _\xi$ to the $\star$-product of two functions
\begin{equation}
X^\star _\xi\triangleright(f\star g) = \left( \xi (f\star g)\right) .\label{5.4}
\end{equation}
To get a better understanding we calculate the right hand side of (\ref{5.4}) to first order in $\theta$ explicitly 
\begin{eqnarray}
(\xi (f\star g)) \hspace*{-2mm}&=&\hspace*{-2mm} \Big(\xi\big( fg +\frac{i}{2}\theta^{\rho\sigma}(\p_\rho f)(\p_\sigma g) \big)\Big) \nonumber\\
&=&\hspace*{-2mm} (\xi f)g + f(\xi g) + \frac{i}{2}\theta^{\rho\sigma}\Big( (\xi\p_\rho f)(\p_\sigma g) 
+ (\p_\rho f)(\xi\p_\sigma g)\Big) +\dots . \label{5.5}
\end{eqnarray}
We have to express the right hand side entirely in terms of operations on ${\cal A}_\theta$
\begin{eqnarray}
(\xi (f\star g)) \hspace*{-2mm}&=&\hspace*{-2mm} (\xi f)\star g + f\star (\xi g) \nonumber\\
&&-\frac{i}{2}\theta^{\rho\sigma}\Big( (\p_\rho (\xi^\mu\p_\mu f))(\p_\sigma g) 
+ (\p_\rho f)(\p_\sigma(\xi^\mu\p_\mu g))\Big) \nonumber\\
&&+ \frac{i}{2}\theta^{\rho\sigma}\Big( (\xi^\mu(\p_\mu\p_\rho f))(\p_\sigma g) 
+ (\p_\rho f)(\xi^\mu(\p_\mu\p_\sigma g)) \Big) + \dots . \label{5.6} 
\end{eqnarray}
Up to first order in $\theta$ this is identical to \cite{Dimitrijevic:2004rf}
\begin{eqnarray}
X^\star _\xi \triangleright (f\star g) \hspace*{-2mm}&=&\hspace*{-2mm} (X^\star_\xi \star f)\star g 
+ f\star (X^\star_\xi \star g) \nonumber\\
&&-\frac{i}{2}\theta^{\rho\sigma}\Big( \lb \p_\rho, \xi^\mu\rb (\p_\mu f)(\p_\sigma g) 
+ (\p_\rho f) \lb \p_\sigma ,\xi^\mu \rb (\p_\mu g)\Big) \nonumber\\
&=&\hspace*{-2mm} (X^\star_\xi \star f)\star g 
+ f\star (X^\star_\xi \star g) \label{5.7}\\
&&-\frac{i}{2}\theta^{\rho\sigma}\Big( (\lb \p^\star_\rho \ds X^\star_\xi\rb \star f)\star (\p^\star_\sigma\star g) + (\p^\star_\rho\star f) \star (\lb \p^\star _\sigma \ds X^\star_\xi\rb \star g)\Big) .\nonumber
\end{eqnarray}
This Leibniz rule follows from an abstract comultiplication rule which reads up to first order in $\theta$
\begin{eqnarray}
\Delta(X^\star_\xi) (f\otimes g) \hspace*{-2mm}&=&\hspace*{-2mm} (X^\star_\xi \star f)\otimes g 
+ f\star (X^\star_\xi \otimes g) \nonumber\\
&-&\hspace{-0.4cm} \frac{i}{2}\theta^{\rho\sigma}\Big( (X^\star_{\lb \p^\star_\rho \ds \xi \rb} \star f)\otimes (\p^\star_\sigma\star g) 
+ (\p^\star_\rho\star f)\otimes (X^\star_{\lb \p^\star _\sigma \ds \xi \rb} \star g)\Big). \label{5.8}
\end{eqnarray}
This comultiplication rule differs from the one we obtained for the classical diffeomorphisms. 
These two Hopf algebras are different although they are the same on the algebra level.

The Leibniz rule (\ref{5.7}) can be calculated to all orders in $\theta$, the result is
\begin{equation}
X^\star_\xi \triangleright (f\star g) = \mu_\star \{ e^{-\frac{i}{2}\theta^{\rho\sigma}\p^{\star}_{\rho} \otimes\p^{\star}_{\sigma}}
\Big( X^\star_\xi\otimes 1 + 1\otimes X^\star_\xi\Big)
e^{\frac{i}{2}\theta^{\rho\sigma}\p^{\star}_{\rho} \otimes \p^{\star}_{\sigma}} \triangleright (f\otimes g) \}. \label{5.9}
\end{equation}
The map $\mu_\star$ was defined in (\ref{3.17}). Expanding (\ref{5.9}) to first order in $\theta$ we obtain (\ref{5.7}). In (\ref{5.9}) appear $\star$-commutators of $\star$-derivatives and the operator $ X^\star_\xi $. A short calculation using the explicit expression for $X_\xi^\star$ given in (\ref{3.23}) yields the following equation: 
\begin{equation}\label{star-commutator}
[\partial_\rho^\star\otimes \partial^\star_\sigma \stackrel{\star}{,} X^\star_\xi \otimes 1 ] = [\partial_\rho^\star \stackrel{\star}{,} X^\star_\xi] \otimes \partial^\star_\sigma = X^\star_{(\partial_\rho \xi)} \otimes \partial^\star_\sigma .
\end{equation}

\newpage

Applying this equation inductively we find an expression where the exponential functions in (\ref{5.9}) are expanded to all orders
\begin{eqnarray}
X^\star_\xi \triangleright (f\star g) \hspace*{-2mm}&=&\hspace*{-2mm} (X^\star_\xi \triangleright f )\star g +  f\star (X^\star_\xi \triangleright g) \nonumber \\ 
&+& \sum_{n=1}^{\infty} \frac{1}{n!}(-\frac{i}{2})^n\theta^{\rho_1 \sigma_1}\cdots \theta^{\rho_n \sigma_n} \Big( (X^\star_{(\partial_{\rho_1} \cdots \partial_{\rho_n} \xi)} \triangleright f ) \star (\p_{\sigma_1} \cdots \p_{\sigma_n} g) \nonumber \\
&  & \qquad + (\p_{\rho_1} \cdots \p_{\rho_n} f) \star (X^\star_{(\partial_{\sigma_1} \cdots \partial_{\sigma_n} \xi)} \triangleright g ) \Big). 
\end{eqnarray}  
Note that $(\partial_\rho \xi )$ and all higher order derivatives of $\xi$ are vector fields again.
    
We outline the calculation leading to (\ref{5.9}) and start from (\ref{5.4})
\begin{equation}
(\xi(f\star g)) =\xi \mu \{ e^{\frac{i}{2}\theta^{\rho\sigma}\p_\rho \otimes\p_\sigma} f\otimes g \} .\label{5.10}
\end{equation}
To commute $\xi$ with $\mu$ we use
\begin{equation}
\xi \mu = \mu \{ \xi\otimes 1+ 1\otimes \xi \}, \label{5.11}
\end{equation}
which can be verified directly by applying it to the tensor product of two functions. We obtain from (\ref{5.10})
\begin{eqnarray}
(\xi(f\star g)) \hspace*{-2mm}&=&\hspace*{-2mm} \mu \{ (\xi\otimes 1+ 1\otimes \xi) \nonumber\\
&&\sum_n \frac{1}{n!}\Big(-\frac{i}{2}\Big)^n
 \theta^{\rho_1\sigma_1}\dots\theta^{\rho_n\sigma_n}(\partial_{\rho_1}\dots\partial_{\rho_n}f)\otimes
 (\partial _{\sigma_1}\dots\partial _{\sigma_n}g) \}.\label{5.12}
\end{eqnarray}
Next we use the fact that $\xi$ applied to derivatives of a function can be mapped to ${\cal A}_\theta$ as in (\ref{5.2}) because derivatives of functions are functions again. This way we express everything in terms of operators defined on ${\cal A}_\theta$. Now we follow the step outlined in (\ref{5.7}) and obtain the result (\ref{5.9}).

Let us summarize the Hopf algebra structure of the diffeomorphism algebra on ${\cal A}_\theta$: For an element $f$ of ${\cal A}_\theta$ we define the transformation
\begin{equation}
\delta_\xi f = -X^\star_\xi \triangleright f \equiv \hat{\delta}_\xi f .\label{5.13}
\end{equation}
This can be used to define $\hat{\delta}_\xi $ as an abstract element of an algebra independent of
its representation as a differential operator. From (\ref{5.3}) the defining relation of the algebra follows
\begin{equation}
\lb \hat{\delta}_\xi , \hat{\delta}_\eta \rb = \hat{\delta}_{\lb \xi,\eta\rb }=\hat{\delta}_{\xi
\times \eta } ,\label{5.14}
\end{equation}
where $\xi$ and $\eta$ are vector fields and $\lb \xi,\eta\rb $ their commutator. The comultiplication from which the Leibniz rule (\ref{5.9}) follows is\footnote{The derivative $\p ^\star_\rho$ can be considered as a variation 
$\hat{\delta}_{\p_\rho} = -\p^\star_\rho$ in the direction of $\p_\rho$.}
\begin{equation}
\Delta(\hat{\delta}_\xi) =  e^{-\frac{i}{2}\theta^{\rho\sigma}\p^{\star}_\rho \otimes\p^{\star}_\sigma}
\Big( \hat{\delta}_\xi\otimes 1 + 1\otimes \hat{\delta}_\xi\Big)
e^{\frac{i}{2}\theta^{\rho\sigma}\p^{\star}_\rho \otimes\p^{\star}_\sigma} . \label{5.15}
\end{equation}
Here the $\star$-commutator of a $\star$-derivative and $\hat{\delta}_\xi$ is given by 
\begin{equation}
[\partial^\star_\rho \stackrel{\star}{,} \hat{\delta}_\xi ] = \hat{\delta}_{(\partial_\rho \xi)} .
\end{equation}
This is the abstract version of (\ref{star-commutator}).
We show that the above comultiplication is compatible with the algebra
\begin{eqnarray}
\lb \Delta(\hat{\delta}_\xi) , \Delta(\hat{\delta}_\eta) \rb \hspace*{-2mm}&=&\hspace*{-2mm}
\lb e^{-\frac{i}{2}\theta^{\rho\sigma}\p^{\star}_\rho \otimes\p^{\star}_\sigma}
\big( \hat{\delta}_\xi\otimes 1 + 1\otimes \hat{\delta}_\xi\big), \big( \hat{\delta}_\eta\otimes 1 + 1\otimes \hat{\delta}_\eta\big)
e^{\frac{i}{2}\theta^{\rho\sigma}\p^{\star}_\rho \otimes\p^{\star}_\sigma} \rb  \nonumber\\
&=&\hspace*{-2mm}e^{-\frac{i}{2}\theta^{\rho\sigma}\p^{\star}_\rho \otimes\p^{\star}_\sigma}
\Big( \hat{\delta}_{\xi\times\eta}\otimes 1 + 1\otimes \hat{\delta}_{\xi\times\eta}\Big)
e^{\frac{i}{2}\theta^{\rho\sigma}\p^{\star}_\rho \otimes\p^{\star}_\sigma} 
= \Delta(\hat{\delta}_{\lb \xi,\eta\rb }) .\label{5.16}
\end{eqnarray}
Coassociativity can be shown as well, counit and antipode can be defined.

Thus, we have obtained a Hopf algebra with the same algebraic relations as for the ordinary diffeomorphism algebra, but the comultiplication is different.

For later use we expand the comultiplication (\ref{5.15}) to first order in $\theta$
\begin{eqnarray}
\Delta(\hat{\delta}_\xi) \hspace*{-2mm}&=&\hspace*{-2mm} \hat{\delta}_\xi\otimes 1 
+ 1\otimes \hat{\delta}_\xi
-\frac{i}{2}\theta^{\rho\sigma}\Big( \lb \p^{\star}_\rho, \hat{\delta}_\xi\rb \otimes \p^{\star}_\sigma + \p^{\star}_\rho \otimes
\lb \p^{\star}_\sigma, \hat{\delta}_\xi\rb \Big) \nonumber\\
&=&\hspace*{-2mm} \hat{\delta}_\xi\otimes 1 + 1\otimes \hat{\delta}_\xi
-\frac{i}{2}\theta^{\rho\sigma}\Big( \hat{\delta}_{(\p_\rho \xi)}\otimes \p^{\star}_\sigma 
+ \p^{\star}_\rho\otimes \hat{\delta}_{(\p_\sigma \xi)}\Big) .\label{5.17}
\end{eqnarray}

\Section{Poincar\' e algebra}

The classical vector fields (\ref{5.1}), when linear in $x$, form a subalgebra of the algebra of diffeomorphisms
\begin{eqnarray}
\xi_\omega \hspace*{-2mm}&=&\hspace*{-2mm} x^\mu\omega_\mu^{\ \nu} \p_\nu , \nonumber\\
\lb \xi_\omega, \xi_{\omega'} \rb \hspace*{-2mm}&=&\hspace*{-2mm} \xi_{\lb \omega, \omega' \rb } , \label{6.1}
\end{eqnarray}
where $\lb \omega, \omega' \rb$ is the commutator of the matrices $\omega$. The corresponding operators $X^\star_\omega$ are
\begin{equation}
X^\star_\omega = x^\mu\omega_\mu^{\ \nu}\p_\nu^\star 
-\frac{i}{2}\theta^{\rho\sigma}\omega_\rho^{\ \nu}\p_\nu^\star \p_\sigma^\star  .\label{6.2}
\end{equation}
Since $\xi_\omega$ is linear in $x$ this is already the exact expression to all orders in $\theta$. The higher order differential operators $X^\star_\omega$  satisfy the same algebra as the vector fields $\xi_\omega$:
\begin{equation}
\lb X_\omega^\star \ds X^\star_{\omega'} \rb = X^\star_{\lb \omega, \omega' \rb} .\label{6.3}
\end{equation}
The transformation defined in (\ref{5.13}) becomes
\begin{equation}
\hat{\delta}_\omega f = -X_\omega^\star \triangleright f = -(\xi_\omega\cdot f) .\label{6.4}
\end{equation}
These transformations together with the derivatives form a Hopf algebra, the relevant algebraic relations follow from (\ref{5.14}) and (\ref{5.15}) and the respective formulas for the derivatives.\\
{\bf Algebra:}
\begin{eqnarray}
\lb \p^\star_\mu , \p^\star_\nu\rb \hspace*{-2mm}&=&\hspace*{-2mm} 0, \nonumber\\
\lb \hat{\delta}_\omega, \hat{\delta}_{\omega'} \rb \hspace*{-2mm}&=&\hspace*{-2mm} \hat{\delta}_{\lb \omega, \omega' \rb},\nonumber\\
\lb \hat{\delta}_\omega, \p^\star_\rho\rb \hspace*{-2mm}&=&\hspace*{-2mm} \omega_\rho^{\ \mu}\p^\star_\mu, \label{6.5}
\end{eqnarray}
{\bf Comultiplication:}
\begin{eqnarray}
\Delta \partial^\star _\mu \hspace*{-2mm}&=&\hspace*{-2mm} \partial^\star _\mu \otimes 1 
+ 1\otimes \partial^\star _\mu,   \nonumber\\
\Delta\hat{\delta}_\omega \hspace*{-2mm}&=&\hspace*{-2mm} \hat{\delta}_\omega\otimes 1 
+  1 \otimes\hat{\delta}_\omega
- \frac{i}{2} \theta^{\rho\sigma}\Big( \lb \p^\star_\rho, \hat{\delta}_\omega\rb \otimes \p^\star_\sigma
+ \p^\star_\rho \otimes \lb \p^\star_\sigma, \hat{\delta}_\omega\rb   \Big)  .\label{6.6}
\end{eqnarray}
This comultiplication mixes the $\hat{\delta}_\omega$ transformations and the derivatives. The transformations (\ref{6.4})
do not form a Hopf algebra by themselves.

We can choose matrices $\omega$ that represent the Lorentz algebra
\begin{equation}
\lb M^{\rho\si},M^{\ka\la} \rb = \eta^{\rho\la}M^{\si\ka}+\eta^{\si\ka}M^{\rho\la}
-\eta^{\rho\ka}M^{\si\la}-\eta^{\si\la}M^{\rho\ka}. \label{6.7}
\end{equation}
With derivatives representing the translations we have obtained a Hopf algebra version of the Poincar\' e algebra
\cite{Oeckl:2000eg,Wess:2003da,Koch:2004ud,Chaichian:2004za,Matlock:2005zn,Calmet:2004ii}. The comultiplication is nontrivially deformed.

The algebra (\ref{6.5}) and (\ref{6.6}) can also be represented by tensor or spinor fields. Let $\psi_A$ be a representation of the Lorentz algebra
\begin{equation}
\hat{\delta}_\omega \psi_A = \omega_\rho^{\ \mu} (M_\mu^{\ \rho})_A^{\ B}\psi_B , \label{6.8}
\end{equation}
where $(M_\mu^{\ \rho})_A^{\ B}$ as a matrix with indices $A$, $B$ represents the Lorentz algebra (\ref{6.7}). The transformation $\hat{\delta}_\omega$ can be defined by the "field transformations"
\begin{equation}
\hat{\delta}_\omega \psi_A = -X^\star_\omega\triangleright \psi_A + \omega_\rho^{\ \mu} (M_\mu^{\ \rho})_A^{\ B}\psi_B . \label{6.9}
\end{equation}
%The derivatives act as usual, e.g. the derivative of a scalar field transforms like a covariant vector field.

With (\ref{6.9}) we have established a Poincar\' e covariant tensor calculus on ${\cal A}_\theta$. The new comultiplication guarantees that the $\star$-product of tensor fields transforms as a tensor. 

Now we can construct Poincar\' e covariant Lagrangians. As an example we discuss a scalar field. Let $\phi$ be a classical scalar field
\begin{equation}
\delta_\omega \phi =-(\xi_\omega\phi) .\label{6.10}
\end{equation}
The transformation law can be mapped to ${\cal A}_\theta$ and we can consider $\phi$ as an element (field) in ${\cal A}_\theta$ with the transformation law
\begin{equation}
\hat{\delta}_\omega \phi = -X^\star_\omega \triangleright \phi .\label{6.11}
\end{equation}
The $\star$-derivative of a scalar field will transform like a vector field
\begin{equation}
\hat{\delta}_\omega (\p_\rho^\star \triangleright\phi) = -X^\star_\omega\triangleright(\p_\rho^\star \triangleright \phi) - X^\star_{(\p_\rho \xi^\mu)}\triangleright (\p^\star_\mu \triangleright \phi) .\label{6.12}
\end{equation}
Thus, the Lagrangian
\begin{equation}
{\cal L} = \frac{1}{2}(\p^\star_\mu \phi)\star(\p^{\star\mu} \phi) -\frac{m^2}{2}\phi\star \phi 
- \lambda \phi\star\phi\star\phi \label{6.13}
\end{equation}
is covariant
\begin{equation}
\hat{\delta}_\omega {\cal L} =-X^\star _{\xi_\omega}\triangleright {\cal L} 
= - \xi_\omega^\lambda(\p_\lambda {\cal L}) .\label{6.14}
\end{equation}
It can be expanded in $\theta$ and to second order we obtain
\begin{eqnarray}
{\cal L} \hspace*{-2mm}&=&\hspace*{-2mm} \frac{1}{2}(\p_\mu \phi)(\p^\mu \phi) -\frac{m^2}{2}\phi\phi 
- \lambda \phi^3  
-\frac{1}{16}\theta^{\rho\sigma}\theta^{\alpha\beta} (\p_\rho\p_\alpha\p_\mu \phi)(\p_\sigma\p_\beta\p^\mu \phi) \nonumber\\
&&+\frac{m^2}{16}\theta^{\rho\sigma}\theta^{\alpha\beta} (\p_\rho\p_\alpha \phi)(\p_\sigma\p_\beta \phi)
+\frac{3}{8}\lambda\theta^{\rho\sigma}\theta^{\alpha\beta} \phi (\p_\rho\p_\alpha \phi)(\p_\sigma\p_\beta \phi)
. \label{6.15}
\end{eqnarray}
Note that a classical transformation of the fields in (\ref{6.15}) will only reproduce (\ref{6.14}) if $\theta$ is transformed as well. Due to the comultiplication rule (\ref{6.6}) we don't have to transform $\theta$ to obtain an invariant action.

To construct an invariant action we define the integration on ${\cal A}_\theta$ as the usual integration. 
This integral has the cyclic property  
\begin{equation}
\int{\mbox{d}}^n x\hspace{1mm} \phi\star\chi = \int{\mbox{d}}^n x\hspace{1mm} \chi\star\phi 
=\int{\mbox{d}}^n x\hspace{1mm} \phi\chi ,\label{6.16}
\end{equation}
which follows by partial integration.
The action
\begin{eqnarray}
S = \int{\mbox{d}}^n x\hspace{1mm} {\cal L} \nonumber
\end{eqnarray}
is invariant if ${\cal L}$ transforms like (\ref{6.14}). 

To derive the equations of motion we vary the action with respect to the field $\phi$. We use the undeformed Leibniz rule for this functional variation and we can use property (\ref{6.16}) to cycle the varied field to the very right (or left) of the integrand. For the Lagrangian (\ref{6.13}) we obtain
\begin{eqnarray}
\delta_\phi S \hspace*{-2mm}&=&\hspace*{-2mm} \delta_\phi \Bigg( \int{\mbox{d}}^n x\hspace{1mm}
\Big( -\frac{1}{2}\phi\star(\p^{\star\mu}\p^\star_\mu \phi) -\frac{m^2}{2}\phi\star\phi 
- \lambda \phi\star\phi\star\phi \Big)\Bigg) \nonumber\\
&=&\hspace*{-2mm}  \int{\mbox{d}}^n x\hspace{1mm} \delta_\phi \phi (x)\star
\Big( -2\frac{1}{2}(\p^{\star\mu}\p^\star_\mu \phi) -2\frac{m^2}{2}\phi 
- 3\lambda \phi\star\phi \Big)  .\label{6.17}
\end{eqnarray}
This leads to the field equations 
\begin{equation}
(\p^{\star\mu}\p_\mu^\star \phi) + m^2\phi + 3\lambda \phi\star\phi =0 .\label{6.18}
\end{equation}
If we expand (\ref{6.18}) in $\theta$ we obtain to second order in $\theta$ the same field equation as from the variation of the action corresponding to the Lagrangian (\ref{6.15})
\begin{equation}
S = \int{\mbox{d}}^n x\hspace{1mm} \Big( \frac{1}{2}(\p_\mu \phi)(\p^\mu \phi) -\frac{m^2}{2}\phi\phi 
- \lambda \phi^3  + \frac{3}{8}\lambda\theta^{\rho\sigma}\theta^{\alpha\beta} \phi (\p_\rho\p_\alpha \phi)(\p_\sigma\p_\beta \phi)  \Big) . \label{6.19}
\end{equation}
Some partial integration is necessary. 
This example will guide us by the construction of a gravity action.

\Section{Differential geometry on ${\cal A}_\theta$}

Gravity theories in general rely on general coordinate transformations which are hard to generalize to noncommutative spaces. The important concept however, on which the gravity theories are based, is the algebra of diffeomorphisms. General relativity can be seen as a theory covariant under diffeomorphisms. We have learned how to deform the diffeomorphism algebra, thus we can construct a deformed gravity theory as a theory covariant under deformed diffeomorphisms.

In section 5 we realized the algebra of vector fields on ${\cal A}_f$ on the noncommutative space ${\cal A}_\theta$. We now develop a tensor calculus for the deformed algebra in analogy to the tensor calculus of the deformed Poincar\' e algebra.

We define the transformation law of a scalar field
\begin{equation}
\hat{\delta}_\xi \phi = -X^\star_\xi\triangleright \phi ,\label{7.1}
\end{equation}
of a covariant vector field
\begin{equation}
\hat{\delta}_\xi V_\mu = -X^\star_\xi \triangleright V_\mu - X^\star_{(\p_\mu \xi^\rho)}\triangleright V_\rho ,\label{7.2}
\end{equation}
of a contravariant vector field
\begin{equation}
\hat{\delta}_\xi V^\mu = -X^\star_\xi \triangleright V^\mu + X^\star_{(\p_\rho \xi^\mu)}\triangleright V^\rho ,\label{7.3}
\end{equation}
and of a general tensor field
\begin{eqnarray}
\hat{\delta}_\xi T_{\mu_1\dots\mu_p}^{\nu_1\dots\nu_r} \hspace*{-2mm}&=&\hspace*{-2mm}
 -X^\star_\xi \triangleright T_{\mu_1\dots\mu_p}^{\nu_1\dots\nu_r}
 - X^\star_{(\p_{\mu_1} \xi^\rho)}\triangleright T_{\rho\dots\mu_p}^{\nu_1\dots\nu_r} -\dots
 - X^\star_{(\p_{\mu_p} \xi^\rho)}\triangleright T_{\mu_1\dots\rho}^{\nu_1\dots\nu_r} \nonumber\\
 && + X^\star_{(\p_\rho \xi^{\nu_1})}\triangleright T_{\mu_1\dots\mu_p}^{\rho\dots\nu_r} +\dots
 + X^\star_{(\p_\rho \xi^\nu_r)}\triangleright T_{\mu_1\dots\mu_p}^{\nu_1\dots\rho}    .\label{7.4}
\end{eqnarray}
The operators $X^\star_\xi$ and $X^\star_{(\p_\mu \xi^\lambda)}$ follow from (\ref{3.23})
\begin{eqnarray}
X^\star_\xi \hspace*{-2mm}&=&\hspace*{-2mm} \sum_{n=0}^\infty \frac{1}{n!}\Big(-\frac{i}{2}\Big)^n
\theta^{\rho_1\sigma_1}\dots\theta^{\rho_n\sigma_n}(\partial_{\rho_1}\dots\partial_{\rho_n}\xi)
\star \partial^\star_{\sigma_1}\dots\partial^\star_{\sigma_n}  \nonumber\\
&=&\hspace*{-2mm}\sum_{n=0}^\infty \frac{1}{n!}\Big(-\frac{i}{2}\Big)^n
\theta^{\rho_1\sigma_1}\dots\theta^{\rho_n\sigma_n}(\partial_{\rho_1}\dots\partial_{\rho_n}\xi^\lambda)
\star \partial^\star_{\sigma_1}\dots\partial^\star_{\sigma_n}\p^\star_\lambda ,\label{7.5} \\
X^\star_{(\p_\mu \xi^\lambda)} \hspace*{-2mm}&=&\hspace*{-2mm}
\sum_{n=0}^\infty \frac{1}{n!}\Big(-\frac{i}{2}\Big)^n
\theta^{\rho_1\sigma_1}\dots\theta^{\rho_n\sigma_n}(\partial_{\rho_1}\dots\partial_{\rho_n}\p_\mu\xi^\lambda)
\star \partial^\star_{\sigma_1}\dots\partial^\star_{\sigma_n}  .\label{7.6}
\end{eqnarray}

The Leibniz rule that follows from (\ref{5.15}) can be defined for the action of $\hat{\delta}_\xi$ on the $\star$-product of any of these fields
\begin{eqnarray}
&&\hat{\delta}_\xi (T_{\mu_1\dots\mu_p}^{\nu_1\dots\nu_r}\star T_{\alpha_1\dots\alpha_s}^{\beta_1\dots\beta_t}) \nonumber\\
&&\hspace*{5mm}= \mu_\star \{ e^{-\frac{i}{2}\theta^{\rho\sigma}\p^\star_\rho \otimes\p^\star_\sigma}
\Big( \hat{\delta}_\xi\otimes 1 + 1\otimes \hat{\delta}_\xi\Big)
e^{\frac{i}{2}\theta^{\rho\sigma}\p^\star_\rho \otimes\p^\star_\sigma}  \triangleright
(T_{\mu_1\dots\mu_p}^{\nu_1\dots\nu_r}\otimes T_{\alpha_1\dots\alpha_s}^{\beta_1\dots\beta_t}) \}. \label{7.7}
\end{eqnarray}
This definition of the comultiplication ensures that the $\star$-product
$T_{\mu_1\dots\mu_p}^{\nu_1\dots\nu_r}\star T_{\alpha_1\dots\alpha_s}^{\beta_1\dots\beta_t}$
transforms like the tensor field $T_{\mu_1\dots\mu_p\alpha_1\dots\alpha_s}^{\nu_1\dots\nu_r\beta_1\dots\beta_t}$.

{\bf Examples:}

The $\star$-product of two scalar fields is a scalar field again
\begin{eqnarray}
\hat{\delta}_\xi(\phi\star\psi) \hspace*{-2mm}&=&\hspace*{-2mm}
\mu_\star \{e^{-\frac{i}{2}\theta^{\gamma\delta}\partial^\star_{\gamma}\otimes\partial^\star_{\delta}}
(\hat{\delta}_\xi\otimes1 + 1\otimes\hat{\delta}_\xi)
e^{\frac{i}{2}\theta^{\rho\sigma}\partial^\star_{\rho}\otimes\partial^\star_{\sigma}}\triangleright (\phi\otimes\psi)\} \nonumber\\
%&=&\hspace*{-2mm} \mu \{e^{\frac{i}{2}\theta^{\alpha\beta}\partial_{\alpha}
%\otimes\partial_{\beta}}e^{-\frac{i}{2}\theta^{\gamma\delta}\partial_{\gamma}\otimes\partial_{\delta}}
%(\hat{\delta}_\xi\otimes1 + 1\otimes\hat{\delta}_\xi)
%e^{\frac{i}{2}\theta^{\rho\sigma}\partial_{\rho}\otimes\partial_{\sigma}}\}\star (\phi\otimes\psi) \nonumber\\
%&=&\hspace*{-2mm}  -\mu \{(X^\star_\xi\otimes 1 + 1\otimes X^\star_\xi)
%e^{\frac{i}{2}\theta^{\rho\sigma}\partial_{\rho}\otimes\partial_{\sigma}} \}\star (\phi\otimes\psi) \nonumber\\
%&=&\hspace*{-2mm} -\mu (\xi\otimes 1 + 1\otimes\xi)
%(e^{\frac{i}{2}\theta^{\rho\sigma}\partial_{\rho}\otimes\partial_{\sigma}}\phi\otimes\psi) \nonumber\\
%&=&\hspace*{-2mm}
%-\xi(\mu (e^{\frac{i}{2}\theta^{\rho\sigma}\partial_{\rho}\otimes\partial_{\sigma}}\phi\otimes\psi)) \nonumber\\
%&=&\hspace*{-2mm} -\xi(\phi\star\psi) = -\xi^\lambda \p_\lambda (\phi\star\psi)  .\label{7.8}
&=&\hspace*{-2mm} -X^\star_\xi\triangleright(\phi\star\psi)  .\label{7.8}
\end{eqnarray}
The proof is the same as for the equation (\ref{5.9}).

Repeating the same calculation one finds that the $\star$-product of a scalar field and a vector field is a vector field
\begin{eqnarray}
\hat{\delta}_\xi(\phi\star V_\mu) = -X^\star_\xi \triangleright (\phi\star V_\mu)
- X^\star_{(\p_\mu \xi^\rho)}\triangleright (\phi\star V_\rho)  , \label{7.9}
\end{eqnarray}
and the $\star$-product of two vector fields is a tensor field
\begin{eqnarray}
\hat{\delta}_\xi(V_\mu\star V_\nu) = -X^\star_\xi \triangleright (V_\mu\star V_\nu)
- X^\star_{(\p_\mu \xi^\rho)}\triangleright (V_\rho\star V_\nu)
- X^\star_{(\p_\nu \xi^\rho)}\triangleright (V_\mu\star V_\rho)  .\label{7.10}
\end{eqnarray}
The contraction of two indices is respected by the transformation law as well:
\begin{eqnarray}
\hat{\delta}_\xi(V_\mu\star V^\mu) \hspace*{-2mm}&=&\hspace*{-2mm}
\mu_\star \{ e^{-\frac{i}{2}\theta^{\rho\sigma}\partial_{\rho}\otimes\partial_{\sigma}}
(\hat{\delta}_{\xi}\otimes1+1\otimes\hat{\delta}_{\xi})
e^{\frac{i}{2}\theta^{\gamma\delta}\partial_{\gamma}\otimes\partial_{\delta}}(V_{\mu}\otimes V^{\mu})\} \nonumber\\
&=&\hspace*{-2mm}  -X^\star_\xi \triangleright (V_{\mu}\star V^{\mu}) .\label{7.11}
\end{eqnarray}
Similar statements are true for the $\star$-product of arbitrary tensors. This is the basic concept of a covariant tensor calculus. Only the derivatives have to be generalized to covariant derivatives by demanding that the covariant derivative itself transforms like a covariant vector
\begin{equation}
\hat{\delta}_{\xi}D_{\mu}V_{\nu} =  -X^\star_\xi \triangleright (D_{\mu}V_{\nu}) - X^\star_{(\p_\mu \xi^\rho)}\triangleright (D_{\rho}V_{\nu})
- X^\star_{(\p_\nu \xi^\rho)}\triangleright (D_{\mu}V_{\rho}) .\label{7.12}
\end{equation}
This can be achieved by introducing a connection $\Gamma_{\mu\nu}^{\alpha}$ and defining the covariant derivative as
\begin{equation}
D_{\mu}V_{\nu} := \partial^\star_{\mu} \triangleright V_{\nu}-\Gamma_{\mu\nu}^{\alpha}\star V_{\alpha}. \label{7.13}
\end{equation}
The transformation law of the connection follows from (\ref{7.12}) if we use the comultiplication (\ref{5.15})
\begin{equation}
\hat{\delta}_{\xi}\Gamma_{\mu\nu}^{\alpha} = -X^\star_\xi \triangleright \Gamma_{\mu\nu}^{\alpha} 
- X^\star_{(\p_\mu \xi^\rho)}\triangleright \Gamma_{\rho\nu}^{\alpha}
- X^\star_{(\p_\nu \xi^\rho)}\triangleright \Gamma_{\mu\rho}^{\alpha} 
+ X^\star_{(\p_\rho \xi^\alpha)}\triangleright \Gamma_{\mu\nu}^{\rho}  
-\partial_{\mu}\partial_{\nu}\xi^{\alpha} .\label{7.14}
\end{equation}
The covariant derivative of a tensor field can be obtained by the same procedure as in the undeformed case, too
\begin{eqnarray}
D_{\lambda}T_{\mu_1\dots\mu_p}^{\nu_1\dots\nu_r} \hspace*{-2mm}&=&\hspace*{-2mm} \partial^\star_{\lambda}\triangleright T_{\mu_1\dots\mu_p}^{\nu_1\dots\nu_r} 
-\Gamma_{\lambda\mu_1}^{\alpha}\star T_{\alpha\dots\mu_p}^{\nu_1\dots\nu_r} -\dots
-\Gamma_{\lambda\mu_p}^{\alpha}\star T_{\mu_1\dots\alpha}^{\nu_1\dots\nu_r} \nonumber\\
&&+\Gamma_{\lambda\alpha}^{\nu_1}\star T_{\mu_1\dots\mu_p}^{\alpha\dots\nu_r} +\dots
+\Gamma_{\lambda\alpha}^{\nu_r}\star T_{\mu_1\dots\mu_p}^{\nu_1\dots\alpha} . \label{7.15}
\end{eqnarray}
Curvature and torsion are obtained in a complete analogy to the undeformed formalism 
\begin{equation}
[D_{\mu} \ds D_{\nu}]\star V_{\rho} = R_{\mu\nu\rho}{}^{\sigma}\star V_{\sigma} + T_{\mu\nu}{}^{\alpha}\star D_{\alpha}V_{\rho} .\label{7.16}
\end{equation}
Using (\ref{7.13}) one finds
\begin{eqnarray}
R_{\mu\nu\rho}{}^{\sigma} \hspace*{-2mm}&=&\hspace*{-2mm} \partial^\star_{\nu}\triangleright\Gamma_{\mu\rho}^{\sigma}
-\partial^\star_{\mu}\triangleright\Gamma_{\nu\rho}^{\sigma}
+\Gamma_{\nu\rho}^{\beta}\star\Gamma_{\mu\beta}^{\sigma}
-\Gamma_{\mu\rho}^{\beta}\star\Gamma_{\nu\beta}^{\sigma}  ,\label{7.17} \\
T_{\mu\nu}{}^{\alpha} \hspace*{-2mm}&=&\hspace*{-2mm} \Gamma_{\nu\mu}^{\alpha}-\Gamma_{\mu\nu}^{\alpha} . \label{7.18}
\end{eqnarray}
For a connection symmetric in $\mu$ and $\nu$ the torsion vanishes.

\Section{Metric and Christoffel symbols}

%Classically, the metric is a symmetric tensor of rang two
%\begin{equation}
%\delta_\xi g_{\mu\nu} = -\xi^\rho(\p_\rho g_{\mu\nu}) - (\p_\mu\xi^\rho) g_{\rho\nu} - 
%(\p_\nu\xi^\rho) g_{\mu\rho} .\label{8.1}
%\end{equation}
%This can be lifted to ${\cal A}_\theta$ by considering $g_{\mu\nu}$ as an element of ${\cal A}_\theta$ and by expressing the transformation properties by operations in ${\cal A}_\theta$
%\begin{equation}
%\hat{\delta}_\xi g_{\mu\nu} = -X^\star _\xi \triangleright g_{\mu\nu} 
%- X^\star _{(\p_\mu\xi^\rho)}\triangleright g_{\rho\nu}
%- X^\star _{(\p_\nu\xi^\rho)} \triangleright g_{\mu\rho} = \delta_\xi g_{\mu\nu} .\label{8.2}
%\end{equation}
%The metric $g_{\mu\nu}$ and its inverse can be used to raise and lower indices.

%In ${\cal A}_\theta$ we have to construct the $\star$-inverse of $g_{\mu\nu}$, we denote it by $g^{\mu\nu\star}$
%\begin{equation}
%g^{\mu\nu\star}\star g_{\nu\rho} = \delta^\mu_\rho . \label{8.3}
%\end{equation}
%The inverse metric $g^{\mu\nu\star}$ is supposed to be a tensor but not a differential operator. To show how such a tensor can be found we first invert a function $f\in {\cal A}_\theta$. As an element of ${\cal A}_f$, $f$ is supposed to have an inverse $f^{-1}$

Classically, the metric is a symmetric tensor of rank two
\begin{equation}
\delta_\xi g_{\mu\nu} = -\xi^\rho(\p_\rho g_{\mu\nu}) - (\p_\mu\xi^\rho) g_{\rho\nu} - 
(\p_\nu\xi^\rho) g_{\mu\rho} .\label{8.1}
\end{equation}
This can be mapped to ${\cal A}_\theta$ by defining $G_{\mu\nu}$ as a symmetric tensor of rank two in 
${\cal A}_\theta$ 
\begin{equation}
\hat{\delta}_\xi G_{\mu\nu} = -X^\star _\xi \triangleright G_{\mu\nu} 
- X^\star _{(\p_\mu\xi^\rho)}\triangleright G_{\rho\nu}
- X^\star _{(\p_\nu\xi^\rho)} \triangleright G_{\mu\rho} ,\label{8.2}
\end{equation}
with the condition that
\begin{equation}
G_{\mu\nu}\Big| _{\theta=0} = g_{\mu\nu} .\label{8.2'}
\end{equation}
A natural choice for $G_{\mu\nu}$ would be $g_{\mu\nu}$ itself. It has the right transformation properties and is $\theta$-independent.

We can also start from four vector fields $E_\mu^{\ a}$ where $\mu$ is the vector index and $a$ numbers the four vector fields. These vector fields can be chosen to be real. The metric can be defined as follows
\begin{equation}
G_{\mu\nu} = \frac{1}{2} \Big( E_\mu^{\ a}\star E_\nu^{\ b} + E_\nu^{\ a}\star E_\mu^{\ b}\Big)\eta_{ab}, \label{8.2''}
\end{equation}
where $\eta_{ab}$ is the $x$-independent symetric metric of flat Minkowski space. With the appropriate comultiplication $G_{\mu\nu}$ is a tensor of second rank. It is symetric by construction and real since $E_\mu^{\ a}$ are real vector fields. To meet condition (\ref{8.2'}) we take the classical vierbein $e_\mu{}^a$ for $E_\mu^{\ a}$. Now 
$G_{\mu\nu}$ is $\theta$-dependent. The metric $G_{\mu\nu}$ and its inverse can be used to raise and lower indices.

In ${\cal A}_\theta$ we have to construct the $\star$-inverse of $G_{\mu\nu}$ which we denote by $G^{\mu\nu\star}$
\begin{equation}
G_{\mu\nu}\star G^{\nu\rho\star}  = \delta_\mu^\rho . \label{8.3}
\end{equation}
The inverse metric $G^{\mu\nu\star}$ is supposed to be a tensor but not a differential operator. To show how such a tensor can be found we first invert a function $f\in {\cal A}_\theta$. As an element of ${\cal A}_f$, $f$ is supposed to have an inverse $f^{-1}$
\begin{equation}
f\cdot f^{-1} =1 .\label{8.4}
\end{equation}
We want to find an inverse of $f$ in ${\cal A}_\theta$, we denote it by $f^{-1\star}$
\begin{equation}
f\star f^{-1\star} =1 .\label{8.5}
\end{equation}
Obviously $f^{-1\star}$ will be different from $f^{-1}$. For its construction we use the geometric
series. We first invert the element
\begin{eqnarray}
f\star f^{-1} =1 + {\cal O}(\theta) ,\label{8.6}
\end{eqnarray}
\begin{eqnarray}
\left( f\star f^{-1}\right) ^{-1\star} \hspace*{-2mm}&=&\hspace*{-2mm}
\left( 1+ f\star f^{-1} -1\right) ^{-1\star} \nonumber\\
&=&\hspace*{-2mm} \sum_{n=0}^\infty \left( 1 - f\star f^{-1} \right) ^{n\star} .\label{8.7}
\end{eqnarray}
The $\star$ on the $n$-th power of $1 - f\star f^{-1}$ indicates that all the products are $\star$-products and therefore
(\ref{8.7}) is an expansion in ${\cal A}_\theta$. Because of (\ref{8.6}) it is also an expansion in $\theta$
\begin{equation}
\left( 1 - f\star f^{-1} \right) ^n = {\cal O}(\theta^n) .\label{8.8}
\end{equation}
From
\begin{equation}
(f\star f^{-1})\star (f\star f^{-1})^{-1\star} =1 \label{8.9}
\end{equation}
and the associativity of the $\star$-product follows
\begin{equation}
f^{-1\star} = f^{-1}\star (f\star f^{-1})^{-1\star} .\label{8.10}
\end{equation}
The $\star$-inverse of $f\star f^{-1}$ has already been calculated as a power series in (\ref{8.7}). Expanding the series (\ref{8.7})
we obtain the following equality which holds up to first order in $\theta$
\begin{eqnarray}
f^{-1\star}\hspace*{-2mm}&=&\hspace*{-2mm} f^{-1} + f^{-1}\star (1 - f\star f^{-1}) \nonumber\\
&=&\hspace*{-2mm} 2f^{-1} - f^{-1}\star f\star f^{-1} , \label{8.11}
\end{eqnarray}
respectively 
\begin{equation}
f^{-1\star}=  3f^{-1} - 3f^{-1}\star f\star f^{-1} + f^{-1}\star f\star f^{-1}\star f\star f^{-1}, \label{8.12}
\end{equation}
which is valid up to second order in $\theta$.
If $f$ transforms classicaly as a scalar field, $f^{-1}$ will transform as a scalar field as well. With the proper comultiplication $f^{-1\star}$ will also be a scalar field.

The same method can be used to find $G^{\mu\nu\star}$
\begin{equation}
G_{\mu\nu}\star G^{\nu\rho\star} = \delta_\mu^\rho . \label{8.13}
\end{equation}
We first invert the matrix
\begin{eqnarray}
G_{\mu\nu}\star G^{\nu\rho} \hspace*{-2mm}&=&\hspace*{-2mm} (G\star G^{-1})^{\ \rho}_\mu = \delta^\rho_\mu
+{\cal O}(\theta) \nonumber\\
{(G\star G^{-1})^{-1\star}} \hspace*{-2mm}&=&\hspace*{-2mm}
\sum_{n=0}^\infty \left( 1 - {G\star G^{-1}}\right)^{n\star} .\label{8.14}
\end{eqnarray}
Here we introduced $G^{\nu\rho}$ as the inverse of $G_{\mu\nu}$ in ${\cal A}_f$,
\begin{eqnarray}
G_{\mu\nu}\cdot G^{\nu\rho} = \delta_\mu^\rho .\nonumber
\end{eqnarray}
For $G_{\mu\nu} = g_{\mu\nu}$, $G^{\mu\nu}$ will be $g^{\mu\nu}$. For $G_{\mu\nu}$ $\theta$-dependent $G^{\mu\nu}$ can be computed by a $\theta$-expansion, starting from $g^{\mu\nu}$ as the $\theta$-independent part.
In analogy to (\ref{8.10}) we obtain
\begin{equation}
G^{\mu\nu\star} = G^{\mu\rho}\star {(G\star G^{-1})^{-1\star}}_\rho^{\ \nu} .\label{8.15}
\end{equation}
When we expand the series $\theta$ we get
\begin{equation}
G^{\mu\nu\star}  = 2G^{\mu\nu} - G^{\mu\alpha}\star G_{\alpha\beta}\star G^{\beta\nu} ,\label{8.16}
\end{equation}
which holds up to first order in $\theta$.
Note that $G^{\mu\nu\star}$ is not a symmetric tensor.

Using the proper coproduct and the fact that $G^{\mu\nu}$ transforms like a contravariant tensor of second rank we conclude that $G^{\mu\nu\star}$ is a tensor of second rank as well
\begin{equation}
\hat{\delta}_\xi G^{\mu\nu\star} = -X^\star _\xi \triangleright G^{\mu\nu\star} 
+ X^\star _{(\p_\rho\xi^\mu)}\triangleright G^{\rho\nu\star}
+ X^\star _{(\p_\rho\xi^\nu)} \triangleright G^{\mu\rho\star} .\label{8.17}
\end{equation}

If we demand that the covariant derivative of the metric vanishes, we can express the symmetric part of the connection entirely in terms of the metric and its derivatives. This is also true in the $\theta$-deformed case.

We shall now assume that the connection is symmetric
\begin{equation}
\Gamma^\rho_{\mu\nu} = \Gamma^\rho_{\nu\mu} ,\label{8.19}
\end{equation}
and when expressed in terms of $G_{\mu\nu}$ we shall call it Christoffel symbol. 

We demand that the covariant derivative of $G_{\mu\nu}$ vanishes
\begin{equation}
D_\alpha G_{\beta\gamma} = \p^\star_\alpha\triangleright G_{\beta\gamma}
- \Gamma_{\alpha\beta}^\rho \star G_{\rho\gamma}- \Gamma_{\alpha\gamma}^\rho \star G_{\beta\rho} =0 . \label{8.20}
\end{equation}
{}From there we proceed as in the classical case. We permute the indices in (\ref{8.20}) assuming from the very beginning
that $G_{\mu\nu}$ is symmetric and add the corresponding equations to obtain
\begin{equation}
2\Gamma^{\rho}_{\alpha\beta}\star G_{\rho\gamma} = \p^\star_\alpha\triangleright G_{\beta\gamma}
+ \p^\star_\beta\triangleright G_{\alpha\gamma} - \p^\star_\gamma\triangleright G_{\alpha\beta} .\label{8.21}
\end{equation}
We can $\star$-invert $G_{\rho\gamma}$ and get the unique result
\begin{equation}
\Gamma^{\sigma}_{\alpha\beta} = \frac{1}{2} \Big( \p^\star_\alpha\triangleright G_{\beta\gamma}
+ \p^\star_\beta\triangleright G_{\alpha\gamma} 
- \p^\star_\gamma\triangleright G_{\alpha\beta}\Big)\star G^{\gamma\sigma\star} .\label{8.22}
\end{equation}
By a direct calculation we can convince ourselves that $\Gamma^{\sigma}_{\alpha\beta}$ has the right transformation property (\ref{7.14}) if $G_{\alpha\beta}$ and $G^{\gamma\sigma\star}$ transform like tensors. All we have used is the symmetry of $G_{\mu\nu}$ and its tensor properties.
%To first order in $\theta$ we obtain
%\begin{equation}
%\Gamma^{\lambda}_{\alpha\beta} = \Gamma^{(0)\lambda}_{\alpha\beta}
%-\frac{i}{2}\theta^{\rho\sigma}(\p_\rho\Gamma^{(0)\mu}_{\alpha\beta})(\p_\sigma g_{\mu\nu})g^{\mu\lambda} ,
%\label{8.23}
%\end{equation}
%where $\Gamma^{(0)\lambda}_{\alpha\beta}$ is the classical expression for the Christoffel symbol
%\begin{equation}
%\Gamma^{(0)\lambda}_{\alpha\beta} = \frac{1}{2} \Big( (\p_\alpha g_{\beta\gamma})
%+ (\p_\beta g_{\alpha\gamma}) - (\p_\gamma g_{\alpha\beta})\Big) g^{\gamma\lambda} .\label{8.24}
%\end{equation}

\Section{Curvature, Ricci tensor and curvature scalar}

To define the curvature tensor we follow the standard procedure. We compute the commutator of two covariant derivatives acting on a vector field. The covariant derivative of a vector field was defined in (\ref{7.13})
\begin{equation}
D_{\mu}V_{\nu} = \partial^\star_{\mu}\triangleright V_{\nu}-\Gamma_{\mu\nu}^{\alpha}\star V_{\alpha}. \label{9.1}
\end{equation}
In (\ref{8.22}) we have found a connection $\Gamma_{\mu\nu}^{\alpha}$ symmetric in $\mu$ and $\nu$
that can be expressed entirely in terms of the metric.
{}From (\ref{7.16}) follows the curvature tensor, because the torsion vanishes for symmetric $\Gamma_{\mu\nu}^{\alpha}$.
\begin{equation}
[D_{\mu} \ds D_{\nu}]\triangleright V_{\rho} = R_{\mu\nu\rho}{}^{\sigma}\star V_{\sigma}. \label{9.3}
\end{equation}
Then the curvature tensor in terms of the Christoffel symbols is given by (\ref{7.17}):
\begin{equation}
R_{\mu\nu\rho}{}^{\sigma} = \partial^\star_{\nu}\triangleright\Gamma_{\mu\rho}^{\sigma}
-\partial^\star_{\mu}\triangleright\Gamma_{\nu\rho}^{\sigma}
+\Gamma_{\nu\rho}^{\beta}\star\Gamma_{\mu\beta}^{\sigma}
-\Gamma_{\mu\rho}^{\beta}\star\Gamma_{\nu\beta}^{\sigma}  .\label{9.4}
\end{equation}
The curvature tensor is antisymmetric in the indices $\mu$ and $\nu$. That the curvature tensor $R_{\mu\nu\rho}{}^{\sigma}$ transform like a tensor if $\Gamma_{\mu\rho}^{\sigma}$ has the transformation property
(\ref{7.14}) can be checked explicitly. Finally, we express the Christoffel symbols in terms of the metric and we obtain the desired form of the curvature tensor in terms of the metric. Its tensor properties then follow from the tensor property of $G_{\mu\nu}$.

{}From the curvature tensor we obtain the Ricci tensor
\begin{equation}
R_{\mu\nu} = R_{\mu\sigma\nu}{}^{\sigma} .\label{9.5}
\end{equation}
A summation over the third index would not vanish as in the undeformed case, but it would not reproduce the Ricci tensor in the limit $\theta\to 0$ either.
The curvature scalar can be defined by contracting the two indices of the Ricci tensor with $G^{\mu\nu\star}$
\begin{equation}
R = G^{\mu\nu\star}\star R_{\nu \mu} .\label{9.6}
\end{equation}
By construction, $R$ transforms as a scalar
\begin{equation}
\hat{\delta}_\xi R = -X^\star_\xi \triangleright R = -\xi^\mu (\p_\mu R) .\label{9.7}
\end{equation}
It will however not be real, as can be seen from (\ref{2.3}). For the Lagrangian to be constructed in the following, we will just add the complex conjugate.

To obtain a covariant action from a scalar that transforms like (\ref{9.7}) we have to find a measure $E$ that transforms as
\begin{equation}
\delta_\xi E = -(\p_\mu (\xi^\mu E)) = -(\p_\mu\xi^\mu) E -\xi^\mu(\p_\mu E) .\label{9.8}
\end{equation}
This has to be mapped to ${\cal A}_\theta$
\begin{equation}
\hat{\delta}_\xi E^\star = -X^\star_\xi\triangleright E^\star 
- X^\star_{(\p_\mu\xi^\mu)}\triangleright E^\star .\label{9.9}
\end{equation}
Such an object we call a scalar $\star$-density. 

Using the comultiplication rule (\ref{5.15}) we can then verify that
\begin{eqnarray}
\hat{\delta}_\xi (E^\star\star R) \hspace*{-2mm}&=&\hspace*{-2mm} -(\p_\mu\xi^\mu) (E^\star\star R) -\xi^\mu(\p_\mu (E^\star\star R)) \nonumber\\
&=&\hspace*{-2mm} - \p_\mu\big(\xi^\mu(E^\star\star R) \big) ,\label{9.10}
\end{eqnarray}
or in the language of ${\cal A}_\theta$
\begin{equation}
\hat{\delta}_\xi (E^\star\star R) = -\p_\mu^\star\triangleright \Big( X^\star_{\xi^\mu}\triangleright (E^\star\star R)\Big) .\label{9.11}
\end{equation}
The action 
\begin{equation}
S = \int{\mbox{d}}^n x\hspace{1mm} E^\star\star R \label{9.12}
\end{equation}
will be invariant
\begin{equation}
\hat{\delta}_\xi \Big( \int{\mbox{d}}^n x\hspace{1mm} E^\star\star R\Big) = 0 .\label{9.13}
\end{equation}

In ${\cal A}_f$ the square root of the determinant of the metric will have the transformation properties of a scalar density. It is however complicated to map the concept of a square root to ${\cal A}_\theta$. It is easier to express the metric in terms of the vierbein as we have done in (\ref{8.2''}) and then define the $\star$-determinant.

The $\star$-determinant can be defined as
\begin{equation}
E^\star = {\mbox {det}}_\star E_\mu^{\ a} = \frac{1}{4!}\varepsilon^{\mu_1\dots\mu_4}\varepsilon_{a_1\dots a_4}
E_{\mu_1}^{\ a_1}\star \dots\star E_{\mu_4}^{\ a_4} .\label{9.15}
\end{equation}
Here we have assumed that our space is four dimensional. The generalization to $n$ dimensions is obvious. 

With this definition $E^\star$ has the right properties of a scalar $\star$-density. To verify this we have to use the transformation properties of covariant vector fields and the comultiplication rule (\ref{5.15}). This reproduces (\ref{9.9}). From the definition also follows that $E^\star$ is real if the vierbeins are real.

An invariant action on ${\cal A}_\theta$ will be
\begin{equation}
S_{\mbox{\tiny EH}} = \frac{1}{2}\int{\mbox{d}}^4 x\hspace{1mm} \Big( E^\star\star R + {\mbox{ c.c.}}\Big) .\label{9.16}
\end{equation}
Using the reality of $E^\star$ and using property  (\ref{6.16}) of the integral we obtain for the action (\ref{9.16}) 
\begin{eqnarray}
S_{\mbox{\tiny EH}} \hspace*{-2mm} &=&\hspace*{-2mm}
\frac{1}{2}\int{\mbox{d}}^4 x\hspace{1mm} E^\star \star (R+\bar{R}) = \frac{1}{2}\int{\mbox{d}}^4 x\hspace{1mm} E^\star  (R+\bar{R}). \label{9.17}
%&=&\hspace*{-2mm}S_{EH}^{(0)}
%+ \frac{1}{2}\int{\mbox{d}}^4 x\hspace{1mm}
%\Big( {\mbox{det}}(e_{\mu}^{\ a})R^{(2)} + E^{\star (2)}R^{(0)} \Big) ,\label{9.17}
\end{eqnarray}
This is the Einstein-Hilbert action on the $\theta$-deformed coordinate space. The field equations can be obtained from this action in analogy to (\ref{6.17}) by moving the field to be varied to the left (or the right) and then varying it, or we could expand the $\star$-products in (\ref{9.17}) and vary the field 
$e_\mu^{\ a}$.

\Section{Expansion in $\theta$}

To get a better insight into the developed formalism it is useful to study a $\theta$-expansion. Already for the gauge theories we used such an expansion for the action and considered $\theta$ as coupling constant. Let us therefore list the $\theta$-expansions of all relevant  quantities. In zeroth order we obtain the classical expressions. We denote them with the index $(0)$.

The basic quantity is the vierbein
\begin{equation}
E_\mu^{\ a} = e_\mu^{\ a} \label{10.1}
\end{equation}
to all orders in $\theta$.

For the metric we obtain
\begin{eqnarray}
G_{\mu\nu} \hspace*{-2mm}&=&\hspace*{-2mm} \frac{1}{2} \Big( E_\mu^{\ a}\star E_\nu^{\ b} + E_\nu^{\ a}\star E_\mu^{\ b}\Big)\eta_{ab}, \nonumber\\
G_{\mu\nu}^{(0)} \hspace*{-2mm}&=&\hspace*{-2mm} e_\mu{}^{a}e_\nu{}^{b}\eta_{ab} = g_{\mu\nu} \label{10.2}
\end{eqnarray}
and up to second order
\begin{equation}
G_{\mu\nu} = g_{\mu\nu}
- \frac{1}{8}\theta^{\alpha_{1}\beta_{1}}\theta^{\alpha_{2}\beta_{2}}
(\partial_{\alpha_{1}}\partial_{\alpha_{2}}e_{\mu}^{\ a})(\partial_{\beta_{1}}\partial_{\beta_{2}}e_{\nu}^{\ b})
\eta_{ab} + \dots.\label{10.3}
\end{equation}
There is no contribution in the first order of $\theta$. The reason is that $\theta$ enters through the $\star$-product only. By definition $G_{\mu\nu}$ is real but the first order in the $\star$-product of two real functions is purely imaginary. Therefore the first order has to vanish and the same will be true for all odd orders in $\theta$.

For $G^{\mu\nu\star}$ we obtain
\begin{eqnarray}
G^{\mu\nu\star} \hspace*{-2mm}&=&\hspace*{-2mm} g^{\mu\nu}-\frac{i}{2}\theta^{\alpha\beta}(\partial_{\alpha}g^{\mu\gamma})
(\partial_{\beta}g_{\gamma\delta})g^{\delta\nu} \nonumber\\
&&+ \frac{1}{8}\theta^{\alpha_{1}\beta_{1}}\theta^{\alpha_{2}\beta_{2}}
\bigg( (\partial_{\alpha_{1}}\partial_{\alpha_{2}}g^{\mu\gamma})(\partial_{\beta_{1}}\partial_{\beta_{2}}g_{\gamma\eta})
+ g^{\mu\gamma}(\partial_{\alpha_{1}}\partial_{\alpha_{2}}e_{\gamma}{}^{a})
(\partial_{\beta_{1}}\partial_{\beta_{2}}e_{\eta}{}^{b})\eta_{ab} \nonumber\\
&&\hspace*{3cm}-2\partial_{\alpha_{1}}\left((\partial_{\alpha_{2}}g^{\mu\gamma})(\partial_{\beta_{2}}g_{\gamma\delta})
 g^{\delta\epsilon}\right)(\partial_{\beta_{1}}g_{\epsilon\eta})\bigg) g^{\eta\nu} .\label{10.4}
\end{eqnarray}
As constructed, $G^{\mu\nu\star}$ is neither symmetric nor real. There is no reason for the term of first order in $\theta$ to drop out. The same is true for the Christoffel symbol and the curvature tensor.
For the Christoffel symbol we get the following expressions up to second order in $\theta$:
The zeroths order reads
\begin{eqnarray}
\Gamma_{\mu\nu}^{(0)\rho} \hspace*{-2mm}&=&\hspace*{-2mm} 
\frac{1}{2} \big( \p_\mu g_{\nu\gamma}
+ \p_\nu g_{\mu\gamma} 
- \p_\gamma g_{\mu\nu}\big) g^{\gamma\rho} ,\label{10.5}
\end{eqnarray}
the first order
\begin{equation}
\Gamma^{(1)}_{\mu\nu}{}^{\rho} = \frac{i}{2}\theta^{\alpha\beta}(\partial_{\alpha}\Gamma_{\mu\nu}^{(0)\sigma})g_{\sigma\tau}
(\partial_{\beta}g^{\tau\rho})
\end{equation}
and the second order
\begin{eqnarray}
\Gamma^{(2)}_{\mu\nu}{}^{\rho} =  
&=&\hspace*{-2mm} - \frac{1}{8}\theta^{\alpha_{1}\beta_{1}}\theta^{\alpha_{2}\beta_{2}}\bigg( (\partial_{\alpha_{1}}\partial_{\alpha_{2}}\Gamma_{\mu\nu\sigma}^{(0)})
(\partial_{\beta_{1}}\partial_{\beta_{2}}g^{\sigma\rho})
-2(\partial_{\alpha_{1}}\Gamma_{\mu\nu\sigma}^{(0)})\partial_{\beta_{1}}\big((\partial_{\alpha_{2}}g^{\sigma\tau})(\partial_{\beta_{2}}g_{\tau\xi})g^{\xi\rho}\big) \nonumber\\
&-&\hspace*{-2mm}\Gamma_{\mu\nu\sigma}^{(0)}\Big(
(\partial_{\alpha_{1}}\partial_{\alpha_{2}}g^{\sigma\tau})(\partial_{\beta_{1}}\partial_{\beta_{2}}g_{\tau\xi})
+g^{\sigma\tau}(\partial_{\alpha_{1}}\partial_{\alpha_{2}}e_{\tau}{}^{a})
(\partial_{\beta_{1}}\partial_{\beta_{2}}e_{\xi}{}^{b})\eta_{ab} \nonumber\\
&-&\hspace*{-2mm}2\partial_{\alpha_{1}}\big( (\partial_{\alpha_{2}}g^{\sigma\tau})(\partial_{\beta_{2}}g_{\tau\lambda})g^{\lambda\kappa}\big)
(\partial_{\beta_{1}}g_{\kappa\xi}) \Big) g^{\xi\rho} 
+\frac{1}{2}\Big( \partial_{\mu}\big( (\partial_{\alpha_{1}}\partial_{\alpha_{2}}e_{\nu}^{\ a}) (\partial_{\beta_{1}}\partial_{\beta_{2}}e_{\sigma}^{\ b})\big) \nonumber\\
&+&\hspace*{-2mm}\partial_{\nu}\big( (\partial_{\alpha_{1}}\partial_{\alpha_{2}}e_{\sigma}^{\ a}) (\partial_{\beta_{1}}\partial_{\beta_{2}}e_{\mu}^{\ b})\big) -\partial_{\sigma}\big( (\partial_{\alpha_{1}}\partial_{\alpha_{2}}e_{\mu}^{\ a}) (\partial_{\beta_{1}}\partial_{\beta_{2}}e_{\nu}^{\ b}) \big) \Big) \eta_{ab}g^{\sigma\rho}\bigg), \label{10.6}
\end{eqnarray}
where
\begin{equation}
\Gamma_{\mu\nu\sigma}^{(0)} = \Gamma_{\mu\nu}^{(0)\rho}g_{\rho\sigma} .
\end{equation}
For the curvature tensor we also list the first and second order individually. The zeroth order is just the classical tensor expressed in the metric or the vierbein
\begin{eqnarray}
R_{\mu\nu\rho}^{(1)}{}^{\sigma} \hspace*{-2mm}&=&\hspace*{-2mm} -\frac{i}{2}\theta^{\kappa\lambda}\bigg( (\partial_{\kappa}R_{\mu\nu\rho}^{(0)}{}^{\tau})
(\partial_{\lambda}g_{\tau\gamma})g^{\gamma\sigma} 
-(\partial_{\kappa}\Gamma_{\nu\rho}^{(0)}{}^{\beta})\Big(
\Gamma_{\mu\beta}^{(0)}{}^{\tau}(\partial_{\lambda}g_{\tau\gamma})g^{\gamma\sigma} \nonumber\\
&&-\Gamma_{\mu\tau}^{(0)}{}^{\sigma}(\partial_{\lambda}g_{\beta\gamma})g^{\gamma\tau}
+\partial_{\mu}\big( (\partial_{\lambda}g_{\beta\gamma})g^{\gamma\sigma}\big)
+(\partial_{\lambda}\Gamma_{\mu\beta}^{(0)}{}^{\sigma})\Big) \nonumber\\
&& +(\partial_{\kappa}\Gamma_{\mu\rho}^{(0)}{}^{\beta})\Big(
\Gamma_{\nu\beta}^{(0)}{}^{\tau}(\partial_{\lambda}g_{\tau\gamma})g^{\gamma\sigma}
-\Gamma_{\nu\tau}^{(0)}{}^{\sigma}(\partial_{\lambda}g_{\beta\gamma})g^{\gamma\tau} \nonumber\\
&&+\partial_{\nu}\big( (\partial_{\lambda}g_{\beta\gamma})g^{\gamma\sigma}\big)
+(\partial_{\lambda}\Gamma_{\nu\beta}^{(0)}{}^{\sigma})\Big) \bigg) \label{10.7}\\
R_{\mu\nu\rho}^{(2)}{}^{\sigma} \hspace*{-2mm}&=&\hspace*{-2mm} \partial_{\nu}\Gamma_{\mu\rho}^{(2)}{}^{\sigma}
+\Gamma_{\nu\rho}^{(2)}{}^{\gamma}\Gamma_{\mu\gamma}^{(0)}{}^{\sigma}
+\Gamma_{\nu\rho}^{(0)}{}^{\gamma}\Gamma_{\mu\gamma}^{(2)}{}^{\sigma} \nonumber\\
&& + \frac{i}{2}\theta^{\alpha\beta}\Big( (\partial_{\alpha}\Gamma_{\nu\rho}^{(1)}{}^{\gamma})(\partial_{\beta}\Gamma_{\mu\gamma}^{(0)}{}^{\sigma})
+(\partial_{\alpha}\Gamma_{\nu\rho}^{(0)}{}^{\gamma})(\partial_{\beta}\Gamma_{\mu\gamma}^{(1)}{}^{\sigma})
\Bigr) \nonumber\\
&& - \frac{1}{8}\theta^{\alpha_{1}\beta_{1}}\theta^{\alpha_{2}\beta_{2}}
(\partial_{\alpha_{1}}\partial_{\alpha_{2}} \Gamma_{\nu\rho}^{(0)}{}^{\gamma})
(\partial_{\beta_{1}}\partial_{\beta_{2}}\Gamma_{\mu\gamma}^{(0)}{}^{\sigma})
\quad -(\mu\leftrightarrow\nu)  .\label{10.8}
\end{eqnarray}
{}From the curvature tensor we obtain the Ricci tensor and the curvature scalar as outlined in the previous section.

The curvature scalar is given by
\begin{equation}
R = R^{(0)} + R^{(1)} + R^{(2)}, \label{10.9}
\end{equation}
where $R^{(0)}$ is the classical curvature scalar and
\begin{eqnarray}
 R^{(1)} \hspace*{-2mm}&=&\hspace*{-2mm} + \frac{i}{2}\theta^{\kappa\lambda}
 \bigg( (\partial_{\kappa}g^{\mu\nu})(\partial_{\lambda}R_{\mu\nu}^{(0)}) 
 -g^{\mu\nu}\Big((\partial_{\kappa}R_{\mu\alpha\nu}^{(0)}{}^{\tau}) 
 (\partial_{\lambda}g_{\tau\gamma})g^{\gamma\alpha} \nonumber\\
 && -(\partial_{\kappa}\Gamma_{\alpha\nu}^{(0)}{}^{\beta})
 \big(\Gamma_{\mu\beta}^{(0)}{}^{\tau}(\partial_{\lambda}g_{\tau\gamma})g^{\gamma\alpha}
 -\Gamma_{\mu\tau}^{(0)}{}^{\alpha}(\partial_{\lambda}g_{\beta\gamma})g^{\gamma\tau}
 +\partial_{\mu}\big((\partial_{\lambda}g_{\beta\gamma})g^{\gamma\alpha}\big) \nonumber\\
 &&+(\partial_{\lambda}\Gamma_{\mu\beta}^{(0)}{}^{\sigma})\big) 
 +(\partial_{\kappa}\Gamma_{\mu\nu}^{(0)}{}^{\beta}) 
 \big(\Gamma_{\alpha\beta}^{(0)}{}^{\tau}(\partial_{\lambda}g_{\tau\gamma})g^{\gamma\alpha}
 -\Gamma_{\nu\tau}^{(0)}{}^{\alpha}(\partial_{\lambda}g_{\beta\gamma})g^{\gamma\tau} \nonumber\\
 &&+\partial_{\alpha}\big((\partial_{\lambda}g_{\beta\gamma})g^{\gamma\alpha}\big)
 +(\partial_{\lambda}\Gamma_{\nu\beta}^{(0)}{}^{\alpha})\big) \Big) \bigg) ,\label{10.10}\\
 R^{(2)} \hspace*{-2mm}&=&\hspace*{-2mm} 
 G^{(2)\mu\nu\star}R_{\nu\mu}^{(0)} + g^{\mu\nu}R_{\nu\mu}^{(2)} + G^{(1)\mu\nu\star}R_{\nu\mu}^{(1)} 
 \nonumber\\
 && +\frac{i}{2}\theta^{\alpha\beta}(\partial_{\alpha}g^{\mu\nu})(\partial_{\beta}R_{\mu\nu}^{(1)})
 -\frac{1}{8}\theta^{\alpha_{1}\beta_{1}}\theta^{\alpha_{2}\beta_{2}}
 (\partial_{\alpha_{1}}\partial_{\alpha_{2}}g^{\mu\nu})(\partial_{\beta_{1}}
 \partial_{\beta_{2}}R_{\mu\nu}^{(0)}). \label{10.11}
 \end{eqnarray}

For the action we still need the scalar $\star$-density $E^\star$
\begin{eqnarray}
E^\star\hspace*{-2mm}&=&\hspace*{-2mm}
{\mbox{det}}(e_\mu^{\ a})
- \frac{1}{8}\frac{1}{4!}\theta^{\alpha_{1}\beta_{1}}\theta^{\alpha_{2}\beta_{2}}
\varepsilon^{\mu_{1}\dots\mu_{4}}\varepsilon_{a_{1}\dots a_{4}} \nonumber\\
&&\Big( (\partial_{\alpha_{1}}\partial_{\alpha_{2}}e_{\mu_{1}}{}^{a_{1}})
(\partial_{\beta_{1}}\partial_{\beta_{2}}e_{\mu_{2}}{}^{a_{2}})e_{\mu_{3}}{}^{a_{3}}e_{\mu_{4}}{}^{a_{4}} \nonumber\\
&&+ \,\partial_{\alpha_{1}}\partial_{\alpha_{2}}(e_{\mu_{1}}{}^{a_{1}}e_{\mu_{2}}{}^{a_{2}})
(\partial_{\beta_{1}}\partial_{\beta_{2}}e_{\mu_{3}}{}^{a_{3}})e_{\mu_{4}}{}^{a_{4}} \nonumber\\
&&+\,\partial_{\alpha_{1}}\partial_{\alpha_{2}}(e_{\mu_{1}}{}^{a_{1}}e_{\mu_{2}}{}^{a_{2}}e_{\mu_{3}}{}^{a_{3}})
(\partial_{\beta_{1}}\partial_{\beta_{2}}e_{\mu_{4}}{}^{a_{4}})\Big) .\label{10.12}
\end{eqnarray}

The Einstein-Hilbert action was defined in (\ref{9.16}). It is real by definition. Since $\theta$ enters only via the $\star$-product we expect that all terms corresponding to odd orders in $\theta$ vanish. Up to second order we therefore get
\begin{equation}
S_{EH} = S_{EH}^{(0)}
+ \int{\mbox{d}}^4 x\hspace{1mm}
\Big( {\mbox{det}}(e_{\mu}^{\ a})R^{(2)} + E^{\star (2)}R^{(0)} \Big) .\label{10.13}
\end{equation}
In this action the even order expansion terms in $\theta$ do not vanish. Equation (\ref{10.12}) allows us to study the deviation from gravity theory on a differential manifold.

%\begin{thebibliography}{99}
%\bibitem{1}
%C.~S.~Chu and P.~M.~Ho,
%{\it Noncommutative open string and D-brane},
%Nucl.\ Phys.\ B {\bf 550}, 151 (1999)
%[hep-th/9812219].

%V.~Schomerus,
%{\it D-branes and deformation quantization},
%JHEP {\bf 9906}, 030 (1999)
%[hep-th/9903205].

%\bibitem{3}
%H.~Weyl, {\it Quantenmechanik und Gruppentheorie}, Z. Phys. {\bf 46}, 1 (1927).

%J.~E.~Moyal, {\it Quantum mechanics as a statistical theory}, Proc.~Cambridge Phil.~Soc. {\bf 45}, 99 (1949).

%\bibitem{4}
%J.~Wess and B.~Zumino,
%{\it Covariant differential calculus on the quantum hyperplane},
%Nucl. Phys. Proc. Suppl. {\bf B18}, 3002-312 (1991).

%S. L. Woronowic,
%{\it Differential calculus on compact matrix pseudogroups (Quantum groups)},
%Commun. Math. Phys. {\bf 122} 125-170 (1989)

%\end{thebibliography}

\bibliographystyle{diss}
\bibliography{literature_gravity}

\end{document}